

%
%
%
\def\unredoffs{} \def\redoffs{\voffset=-.31truein\hoffset=-.48truein}
\def\speclscape{}
%
%
%
%
%

\newbox\leftpage \newdimen\fullhsize \newdimen\hstitle \newdimen\hsbody
\tolerance=1000\hfuzz=2pt
\catcode`\@=11 
\ifx\hyperdef\UNd@FiNeD\def\hyperdef#1#2#3#4{#4}\def\hyperref#1#2#3#4{#4}\fi
\def\bigans{b }
\def\answ{b }
%
\ifx\answ\bigans\message{(This will come out unreduced.}
\magnification=1200\unredoffs\baselineskip=16pt plus 2pt minus 1pt
\hsbody=\hsize \hstitle=\hsize 
\else\message{(This will be reduced.} \let\l@r=L
\magnification=1000\baselineskip=16pt plus 2pt minus 1pt \vsize=7truein
\redoffs \hstitle=8truein\hsbody=4.75truein\fullhsize=10truein\hsize=\hsbody
\output={\ifnum\pageno=0 
  \shipout\vbox{\speclscape{\hsize\fullhsize\makeheadline}
    \hbox to \fullhsize{\hfill\pagebody\hfill}}\advancepageno
  \else
  \almostshipout{\leftline{\vbox{\pagebody\makefootline}}}\advancepageno
  \fi}
\def\almostshipout#1{\if L\l@r \count1=1 \message{[\the\count0.\the\count1]}
      \global\setbox\leftpage=#1 \global\let\l@r=R
 \else \count1=2
  \shipout\vbox{\speclscape{\hsize\fullhsize\makeheadline}
      \hbox to\fullhsize{\box\leftpage\hfil#1}}  \global\let\l@r=L\fi}
\fi
%
\newcount\yearltd\yearltd=\year\advance\yearltd by -2000

\def\Title#1#2{\nopagenumbers\abstractfont\hsize=\hstitle\rightline{#1}%
\vskip 1in\centerline{\titlefont #2}\abstractfont\vskip .5in\pageno=0}
\def\Date#1{\vfill\leftline{#1}\tenpoint\supereject\global\hsize=\hsbody%
\footline={\hss\tenrm\hyperdef\hypernoname{page}\folio\folio\hss}}%
%

\def\draftmode{\message{ DRAFTMODE }\def\draftdate{{\rm preliminary draft:
\number\month/\number\day/\number\yearltd\ \ \hourmin}}%
\headline={\hfil\draftdate}\writelabels\baselineskip=20pt plus 2pt minus 2pt
 {\count255=\time\divide\count255 by 60 \xdef\hourmin{\number\count255}
  \multiply\count255 by-60\advance\count255 by\time
  \xdef\hourmin{\hourmin:\ifnum\count255<10 0\fi\the\count255}}}
\def\nolabels{\def\wrlabeL##1{}\def\eqlabeL##1{}\def\reflabeL##1{}}
\def\writelabels{\def\wrlabeL##1{\leavevmode\vadjust{\rlap{\smash%
{\line{{\escapechar=` \hfill\rlap{\sevenrm\hskip.03in\string##1}}}}}}}%
\def\eqlabeL##1{{\escapechar-1\rlap{\sevenrm\hskip.05in\string##1}}}%
\def\reflabeL##1{\noexpand\llap{\noexpand\sevenrm\string\string\string##1}}}
\nolabels
%
\global\newcount\secno \global\secno=0
\global\newcount\meqno \global\meqno=1
\def\s@csym{}
\def\newsec#1{\global\advance\secno by1%
{\toks0{#1}\message{(\the\secno. \the\toks0)}}%
\global\subsecno=0\eqnres@t\let\s@csym\secsym\xdef\secn@m{\the\secno}\noindent
{\bf\hyperdef\hypernoname{section}{\the\secno}{\the\secno.} #1}%
\writetoca{{\string\hyperref{}{section}{\the\secno}{\the\secno.}} {#1}}%
\par\nobreak\medskip\nobreak}
\def\eqnres@t{\xdef\secsym{\the\secno.}\global\meqno=1\bigbreak\bigskip}
\def\sequentialequations{\def\eqnres@t{\bigbreak}}\xdef\secsym{}
\global\newcount\subsecno \global\subsecno=0
\def\subsec#1{\global\advance\subsecno by1%
{\toks0{#1}\message{(\s@csym\the\subsecno. \the\toks0)}}%
\ifnum\lastpenalty>9000\else\bigbreak\fi
\noindent{\it\hyperdef\hypernoname{subsection}{\secn@m.\the\subsecno}%
{\secn@m.\the\subsecno.} #1}\writetoca{\string\quad
{\string\hyperref{}{subsection}{\secn@m.\the\subsecno}{\secn@m.\the\subsecno.}}
{#1}}\par\nobreak\medskip\nobreak}
\def\appendix#1#2{\global\meqno=1\global\subsecno=0\xdef\secsym{\hbox{#1.}}%
\bigbreak\bigskip\noindent{\bf Appendix \hyperdef\hypernoname{appendix}{#1}%
{#1.} #2}{\toks0{(#1. #2)}\message{\the\toks0}}%
\xdef\s@csym{#1.}\xdef\secn@m{#1}%
\writetoca{\string\hyperref{}{appendix}{#1}{Appendix {#1.}} {#2}}%
\par\nobreak\medskip\nobreak}
%
%
\def\checkm@de#1#2{\ifmmode{\def\f@rst##1{##1}\hyperdef\hypernoname{equation}%
{#1}{#2}}\else\hyperref{}{equation}{#1}{#2}\fi}
\def\eqnn#1{\DefWarn#1\xdef #1{(\noexpand\relax\noexpand\checkm@de%
{\s@csym\the\meqno}{\secsym\the\meqno})}%
\wrlabeL#1\writedef{#1\leftbracket#1}\global\advance\meqno by1}
\def\f@rst#1{\c@t#1a\em@ark}\def\c@t#1#2\em@ark{#1}
\def\eqna#1{\DefWarn#1\wrlabeL{#1$\{\}$}%
\xdef #1##1{(\noexpand\relax\noexpand\checkm@de%
{\s@csym\the\meqno\noexpand\f@rst{##1}}{\hbox{$\secsym\the\meqno##1$}})}
\writedef{#1\numbersign1\leftbracket#1{\numbersign1}}\global\advance\meqno by1}
\def\eqn#1#2{\DefWarn#1%
\xdef #1{(\noexpand\hyperref{}{equation}{\s@csym\the\meqno}%
{\secsym\the\meqno})}$$#2\eqno(\hyperdef\hypernoname{equation}%
{\s@csym\the\meqno}{\secsym\the\meqno})\eqlabeL#1$$%
\writedef{#1\leftbracket#1}\global\advance\meqno by1}
\def\xeqn{\expandafter\xe@n}\def\xe@n(#1){#1}
\def\xeqna#1{\expandafter\xe@n#1}
\def\eqns#1{(\e@ns #1{\hbox{}})}
\def\e@ns#1{\ifx\UNd@FiNeD#1\message{eqnlabel \string#1 is undefined.}%
\xdef#1{(?.?)}\fi{\let\hyperref=\relax\xdef\next{#1}}%
\ifx\next\em@rk\def\next{}\else%
\ifx\next#1\xeqn#1\else\def\n@xt{#1}\ifx\n@xt\next#1\else\xeqna#1\fi
\fi\let\next=\e@ns\fi\next}

\def\DefWarn#1{\ifx\UNd@FiNeD#1\else
\immediate\write16{*** WARNING: the label \string#1 is already defined ***}\fi}
%
\newskip\footskip\footskip14pt plus 1pt minus 1pt 
\def\footnotefont{\ninepoint}\def\f@t#1{\footnotefont #1\@foot}
\def\f@@t{\baselineskip\footskip\bgroup\footnotefont\aftergroup\@foot\let\next}
\setbox\strutbox=\hbox{\vrule height9.5pt depth4.5pt width0pt}
\global\newcount\ftno \global\ftno=0
\def\foot{\global\advance\ftno by1\def\foot@rg{\hyperref{}{footnote}%
{\the\ftno}{\the\ftno}\xdef\foot@rg{\noexpand\hyperdef\noexpand\hypernoname%
{footnote}{\the\ftno}{\the\ftno}}}\footnote{$^{\foot@rg}$}}
%
\newwrite\ftfile
\def\footend{\def\foot{\global\advance\ftno by1\chardef\wfile=\ftfile
\hyperref{}{footnote}{\the\ftno}{$^{\the\ftno}$}%
\ifnum\ftno=1\immediate\openout\ftfile=\jobname.fts\fi%
\immediate\write\ftfile{\noexpand\smallskip%
\noexpand\item{\noexpand\hyperdef\noexpand\hypernoname{footnote}
{\the\ftno}{f\the\ftno}:\ }\pctsign}\findarg}%
\def\footatend{\vfill\eject\immediate\closeout\ftfile{\parindent=20pt
\centerline{\bf Footnotes}\nobreak\bigskip\input \jobname.fts }}}
\def\footatend{}
%
%
\global\newcount\refno \global\refno=1
\newwrite\rfile
\def\ref{[\hyperref{}{reference}{\the\refno}{\the\refno}]\nref}
\def\nref#1{\DefWarn#1%
\xdef#1{[\noexpand\hyperref{}{reference}{\the\refno}{\the\refno}]}%
\writedef{#1\leftbracket#1}%
\ifnum\refno=1\immediate\openout\rfile=\jobname.refs\fi
\chardef\wfile=\rfile\immediate\write\rfile{\noexpand\item{[\noexpand\hyperdef%
\noexpand\hypernoname{reference}{\the\refno}{\the\refno}]\ }%
\reflabeL{#1\hskip.31in}\pctsign}\global\advance\refno by1\findarg}
\def\findarg#1#{\begingroup\obeylines\newlinechar=`\^^M\pass@rg}
{\obeylines\gdef\pass@rg#1{\writ@line\relax #1^^M\hbox{}^^M}%
\gdef\writ@line#1^^M{\expandafter\toks0\expandafter{\striprel@x #1}%
\edef\next{\the\toks0}\ifx\next\em@rk\let\next=\endgroup\else\ifx\next\empty%
\else\immediate\write\wfile{\the\toks0}\fi\let\next=\writ@line\fi\next\relax}}
\def\striprel@x#1{} \def\em@rk{\hbox{}}
\def\lref{\begingroup\obeylines\lr@f}
\def\lr@f#1#2{\DefWarn#1\gdef#1{\let#1=\UNd@FiNeD\ref#1{#2}}\endgroup\unskip}

\def\addref#1{\immediate\write\rfile{\noexpand\item{}#1}} 
\def\listrefs{\footatend\vfill\supereject\immediate\closeout\rfile\writestoppt
\baselineskip=\footskip\centerline{{\bf References}}\bigskip{\parindent=20pt%
\frenchspacing\escapechar=` \input \jobname.refs\vfill\eject}\nonfrenchspacing}
\def\startrefs#1{\immediate\openout\rfile=\jobname.refs\refno=#1}
\def\xref{\expandafter\xr@f}\def\xr@f[#1]{#1}
\def\refs#1{\count255=1[\r@fs #1{\hbox{}}]}
\def\r@fs#1{\ifx\UNd@FiNeD#1\message{reflabel \string#1 is undefined.}%
\nref#1{need to supply reference \string#1.}\fi%
\vphantom{\hphantom{#1}}{\let\hyperref=\relax\xdef\next{#1}}%
\ifx\next\em@rk\def\next{}%
\else\ifx\next#1\ifodd\count255\relax\xref#1\count255=0\fi%
\else#1\count255=1\fi\let\next=\r@fs\fi\next}
%

%
\newwrite\ffile\global\newcount\figno \global\figno=1
\def\fig{fig.~\hyperref{}{figure}{\the\figno}{\the\figno}\nfig}
\def\nfig#1{\DefWarn#1%
\xdef#1{fig.~\noexpand\hyperref{}{figure}{\the\figno}{\the\figno}}%
\writedef{#1\leftbracket fig.\noexpand~\xfig#1}%
\ifnum\figno=1\immediate\openout\ffile=\jobname.figs\fi\chardef\wfile=\ffile%
{\let\hyperref=\relax
\immediate\write\ffile{\noexpand\medskip\noexpand\item{Fig.\ %
\noexpand\hyperdef\noexpand\hypernoname{figure}{\the\figno}{\the\figno}. }
\reflabeL{#1\hskip.55in}\pctsign}}\global\advance\figno by1\findarg}
\def\listfigs{\vfill\eject\immediate\closeout\ffile{\parindent40pt
\baselineskip14pt\centerline{{\bf Figure Captions}}\nobreak\medskip
\escapechar=` \input \jobname.figs\vfill\eject}}
\def\xfig{\expandafter\xf@g}\def\xf@g fig.\penalty\@M\ {}
\def\figs#1{figs.~\f@gs #1{\hbox{}}}
\def\f@gs#1{{\let\hyperref=\relax\xdef\next{#1}}\ifx\next\em@rk\def\next{}\else
\ifx\next#1\xfig #1\else#1\fi\let\next=\f@gs\fi\next}
\def\figin{\epsfcheck\figin}\def\figins{\epsfcheck\figins}
\def\epsfcheck{\ifx\epsfbox\UNd@FiNeD
\message{(NO epsf.tex, FIGURES WILL BE IGNORED)}
\gdef\figin##1{\vskip2in}\gdef\figins##1{\hskip.5in}
\else\message{(FIGURES WILL BE INCLUDED)}%
\gdef\figin##1{##1}\gdef\figins##1{##1}\fi}
\def\DefWarn#1{}
\def\figinsert{\goodbreak\midinsert}
\def\ifig#1#2#3{\DefWarn#1\xdef#1{fig.~\noexpand\hyperref{}{figure}%
{\the\figno}{\the\figno}}\writedef{#1\leftbracket fig.\noexpand~\xfig#1}%
\figinsert\figin{\centerline{#3}}\medskip\centerline{\vbox{\baselineskip12pt
\advance\hsize by -1truein\noindent\wrlabeL{#1=#1}\footnotefont%
{\bf Fig.~\hyperdef\hypernoname{figure}{\the\figno}{\the\figno}:} #2}}
\bigskip\endinsert\global\advance\figno by1}
\newwrite\lfile
{\escapechar-1\xdef\pctsign{\string\%}\xdef\leftbracket{\string\{}
\xdef\rightbracket{\string\}}\xdef\numbersign{\string\#}}
\def\writedefs{\immediate\openout\lfile=\jobname.defs \def\writedef##1{%
{\let\hyperref=\relax\let\hyperdef=\relax\let\hypernoname=\relax
 \immediate\write\lfile{\string\def\string##1\rightbracket}}}}%
\def\writestop{\def\writestoppt{\immediate\write\lfile{\string\pageno
 \the\pageno\string\startrefs\leftbracket\the\refno\rightbracket
 \string\def\string\secsym\leftbracket\secsym\rightbracket
 \string\secno\the\secno\string\meqno\the\meqno}\immediate\closeout\lfile}}
\def\writestoppt{}\def\writedef#1{}
\def\seclab#1{\DefWarn#1%
\xdef #1{\noexpand\hyperref{}{section}{\the\secno}{\the\secno}}%
\writedef{#1\leftbracket#1}\wrlabeL{#1=#1}}
\def\subseclab#1{\DefWarn#1%
\xdef #1{\noexpand\hyperref{}{subsection}{\secn@m.\the\subsecno}%
{\secn@m.\the\subsecno}}\writedef{#1\leftbracket#1}\wrlabeL{#1=#1}}
\def\applab#1{\DefWarn#1%
\xdef #1{\noexpand\hyperref{}{appendix}{\secn@m}{\secn@m}}%
\writedef{#1\leftbracket#1}\wrlabeL{#1=#1}}
\newwrite\tfile \def\writetoca#1{}
\def\leaderfill{\leaders\hbox to 1em{\hss.\hss}\hfill}
\def\writetoc{\immediate\openout\tfile=\jobname.toc
   \def\writetoca##1{{\edef\next{\write\tfile{\noindent ##1
   \string\leaderfill {\string\hyperref{}{page}{\noexpand\number\pageno}%
                       {\noexpand\number\pageno}} \par}}\next}}}
\newread\ch@ckfile
\def\listtoc{\immediate\closeout\tfile\immediate\openin\ch@ckfile=\jobname.toc
\ifeof\ch@ckfile\message{no file \jobname.toc, no table of contents this pass}%
\else\closein\ch@ckfile\centerline{\bf Contents}\nobreak\medskip%
{\baselineskip=12pt\footnotefont\parskip=0pt\catcode`\@=11\input\jobname.toc
\catcode`\@=12\bigbreak\bigskip}\fi}
\catcode`\@=12 
%
\edef\tfontsize{\ifx\answ\bigans scaled\magstep3\else scaled\magstep4\fi}
\font\titlerm=cmr10 \tfontsize \font\titlerms=cmr7 \tfontsize
\font\titlermss=cmr5 \tfontsize \font\titlei=cmmi10 \tfontsize
\font\titleis=cmmi7 \tfontsize \font\titleiss=cmmi5 \tfontsize
\font\titlesy=cmsy10 \tfontsize \font\titlesys=cmsy7 \tfontsize
\font\titlesyss=cmsy5 \tfontsize \font\titleit=cmti10 \tfontsize
\skewchar\titlei='177 \skewchar\titleis='177 \skewchar\titleiss='177
\skewchar\titlesy='60 \skewchar\titlesys='60 \skewchar\titlesyss='60
\def\titlefont{\def\rm{\fam0\titlerm}
\textfont0=\titlerm \scriptfont0=\titlerms \scriptscriptfont0=\titlermss
\textfont1=\titlei \scriptfont1=\titleis \scriptscriptfont1=\titleiss
\textfont2=\titlesy \scriptfont2=\titlesys \scriptscriptfont2=\titlesyss
\textfont\itfam=\titleit \def\it{\fam\itfam\titleit}\rm}
 \ifx\answ\bigans\else scaled\magstep1\fi
\ifx\answ\bigans\def\abstractfont{\tenpoint}\else
\font\absit=cmti10 scaled \magstep1
\font\abssl=cmsl10 scaled \magstep1
\font\absrm=cmr10 scaled\magstep1 \font\absrms=cmr7 scaled\magstep1
\font\absrmss=cmr5 scaled\magstep1 \font\absi=cmmi10 scaled\magstep1
\font\absis=cmmi7 scaled\magstep1 \font\absiss=cmmi5 scaled\magstep1
\font\abssy=cmsy10 scaled\magstep1 \font\abssys=cmsy7 scaled\magstep1
\font\abssyss=cmsy5 scaled\magstep1 \font\absbf=cmbx10 scaled\magstep1
\skewchar\absi='177 \skewchar\absis='177 \skewchar\absiss='177
\skewchar\abssy='60 \skewchar\abssys='60 \skewchar\abssyss='60
\def\abstractfont{\def\rm{\fam0\absrm}
\textfont0=\absrm \scriptfont0=\absrms \scriptscriptfont0=\absrmss
\textfont1=\absi \scriptfont1=\absis \scriptscriptfont1=\absiss
\textfont2=\abssy \scriptfont2=\abssys \scriptscriptfont2=\abssyss
\textfont\itfam=\absit \def\it{\fam\itfam\absit}\def\footnotefont{\tenpoint}%
\textfont\slfam=\abssl \def\sl{\fam\slfam\abssl}%
\textfont\bffam=\absbf \def\bf{\fam\bffam\absbf}\rm}\fi
\def\tenpoint{\def\rm{\fam0\tenrm}
\textfont0=\tenrm \scriptfont0=\sevenrm \scriptscriptfont0=\fiverm
\textfont1=\teni  \scriptfont1=\seveni  \scriptscriptfont1=\fivei
\textfont2=\tensy \scriptfont2=\sevensy \scriptscriptfont2=\fivesy
\textfont\itfam=\tenit \def\it{\fam\itfam\tenit}\def\footnotefont{\ninepoint}%
\textfont\bffam=\tenbf \def\bf{\fam\bffam\tenbf}\def\sl{\fam\slfam\tensl}\rm}
\font\ninerm=cmr9 \font\sixrm=cmr6 \font\ninei=cmmi9 \font\sixi=cmmi6
\font\ninesy=cmsy9 \font\sixsy=cmsy6 \font\ninebf=cmbx9
\font\nineit=cmti9 \font\ninesl=cmsl9 \skewchar\ninei='177
\skewchar\sixi='177 \skewchar\ninesy='60 \skewchar\sixsy='60
\def\ninepoint{\def\rm{\fam0\ninerm}
\textfont0=\ninerm \scriptfont0=\sixrm \scriptscriptfont0=\fiverm
\textfont1=\ninei \scriptfont1=\sixi \scriptscriptfont1=\fivei
\textfont2=\ninesy \scriptfont2=\sixsy \scriptscriptfont2=\fivesy
\textfont\itfam=\ninei \def\it{\fam\itfam\nineit}\def\sl{\fam\slfam\ninesl}%
\textfont\bffam=\ninebf \def\bf{\fam\bffam\ninebf}\rm}
%
%

\hyphenation{anom-aly anom-alies coun-ter-term coun-ter-terms}
\def\inv{^{\raise.15ex\hbox{${\scriptscriptstyle -}$}\kern-.05em 1}}

\def\Dsl{\,\raise.15ex\hbox{/}\mkern-13.5mu D} 
\def\dsl{\raise.15ex\hbox{/}\kern-.57em\partial}

\def\lspace{\ifx\answ\bigans{}\else\qquad\fi}
\def\lbspace{\ifx\answ\bigans{}\else\hskip-.2in\fi} 
\def\boxeqn#1{\vcenter{\vbox{\hrule\hbox{\vrule\kern3pt\vbox{\kern3pt
	\hbox{${\displaystyle #1}$}\kern3pt}\kern3pt\vrule}\hrule}}}
\def\mbox#1#2{\vcenter{\hrule \hbox{\vrule height#2in
		\kern#1in \vrule} \hrule}}  
%

\def\darr#1{\raise1.5ex\hbox{$\leftrightarrow$}\mkern-16.5mu #1}

\def\roughly#1{\raise.3ex\hbox{$#1$\kern-.75em\lower1ex\hbox{$\sim$}}}

\def\smallfig#1#2#3{\DefWarn#1\xdef#1{fig.~\the\figno}
\writedef{#1\leftbracket fig.\noexpand~\the\figno}%
\figinsert\figin{\centerline{#3}}\medskip\centerline{\vbox{
\baselineskip12pt\advance\hsize by -1truein
\noindent\footnotefont{\bf Fig.~\the\figno:} #2}}
\endinsert\global\advance\figno by1}

\def\bb{
\font\tenmsb=msbm10
\font\sevenmsb=msbm7
\font\fivemsb=msbm5
\textfont1=\tenmsb
\scriptfont1=\sevenmsb
\scriptscriptfont1=\fivemsb
}

\input amssym

%
%
\ifx\pdfoutput\undefined
\input epsf
\def\fig#1{\epsfbox{#1.eps}}
\def\figscale#1#2{\epsfxsize=#2\epsfbox{#1.eps}}
%
%
\else
\def\fig#1{\pdfximage {#1.pdf}\pdfrefximage\pdflastximage}
\def\figscale#1#2{\pdfximage width#2 {#1.pdf}\pdfrefximage\pdflastximage}
\fi

\def\IZ{\relax\ifmmode\mathchoice
{\hbox{\cmss Z\kern-.4em Z}}{\hbox{\cmss Z\kern-.4em Z}} {\lower.9pt\hbox{\cmsss Z\kern-.4em Z}}
{\lower1.2pt\hbox{\cmsss Z\kern-.4em Z}}\else{\cmss Z\kern-.4em Z}\fi}

\newif\ifdraft\draftfalse
\newif\ifinter\interfalse
\ifdraft\draftmode\else\interfalse\fi
\def\journal#1&#2(#3){\unskip, \sl #1\ \bf #2 \rm(19#3) }
\def\andjournal#1&#2(#3){\sl #1~\bf #2 \rm (19#3) }

\def\frac#1#2{{#1\over#2}}

\def\inbar{\,\vrule height1.5ex width.4pt depth0pt}
\def\IC{\relax\hbox{$\inbar\kern-.3em{\rm C}$}}
\def\IR{\relax{\rm I\kern-.18em R}}
\def\IP{\relax{\rm I\kern-.18em P}}
\def\Z{{\bf Z}}

%
%


%
\catcode`\@=11
\def\slash#1{\mathord{\mathpalette\c@ncel{#1}}}
\overfullrule=0pt

\def\underrel#1\over#2{\mathrel{\mathop{\kern\z@#1}\limits_{#2}}}

\catcode`\@=12


%

\def\mod{{\rm mod}}
\def \sinh{{\rm sinh}}


\def\[{[}
\def\]{]}

\def\comment#1{ }

%
\def\draftnote#1{\ifdraft{\baselineskip2ex
                 \vbox{\kern1em\hrule\hbox{\vrule\kern1em\vbox{\kern1ex
                 \noindent \underbar{NOTE}: #1
             \vskip1ex}\kern1em\vrule}\hrule}}\fi}
\def\internote#1{\ifinter{\baselineskip2ex
                 \vbox{\kern1em\hrule\hbox{\vrule\kern1em\vbox{\kern1ex
                 \noindent \underbar{Internal Note}: #1
             \vskip1ex}\kern1em\vrule}\hrule}}\fi}

%
%



%
%
%
%

%

\def\inv{^{-1}}


\def\1{{\ds 1}}

\def\Z{\hbox{$\bb Z$}}

%
\def\draftnote#1{\ifdraft{\baselineskip2ex
                 \vbox{\kern1em\hrule\hbox{\vrule\kern1em\vbox{\kern1ex
                 \noindent \underbar{NOTE}: #1
             \vskip1ex}\kern1em\vrule}\hrule}}\fi}
\def\internote#1{\ifinter{\baselineskip2ex
                 \vbox{\kern1em\hrule\hbox{\vrule\kern1em\vbox{\kern1ex
                 \noindent \underbar{Internal Note}: #1
             \vskip1ex}\kern1em\vrule}\hrule}}\fi}

%
%



%
%
%
%

%

\def\inv{^{-1}}


\def\1{{\ds 1}}

\def\Z{\hbox{$\bb Z$}}

\def\S{\hbox{$\bb S$}}

\newfam\frakfam
\font\teneufm=eufm10
\font\seveneufm=eufm7
\font\fiveeufm=eufm5
\textfont\frakfam=\teneufm
\scriptfont\frakfam=\seveneufm
\scriptscriptfont\frakfam=\fiveeufm

\lref\NiarchosAH{
  V.~Niarchos,
  ``Seiberg dualities and the 3d/4d connection,''
JHEP {\bf 1207}, 075 (2012).
[arXiv:1205.2086 [hep-th]].
}

\lref\AharonyGP{
  O.~Aharony,
  ``IR duality in $d = 3$ $\cal{N}=$ 2 supersymmetric USp$(2N_c)$ and U$(N_c)$ gauge theories,''
Phys.\ Lett.\ B {\bf 404}, 71 (1997).
[hep-th/9703215].
}

\lref\AffleckAS{
  I.~Affleck, J.~A.~Harvey and E.~Witten,
  ``Instantons and (Super)Symmetry Breaking in (2+1)-Dimensions,''
Nucl.\ Phys.\ B {\bf 206}, 413 (1982)..
}

\lref\IntriligatorID{
  K.~A.~Intriligator and N.~Seiberg,
  ``Duality, monopoles, dyons, confinement and oblique confinement in supersymmetric SO(N(c)) gauge theories,''
Nucl.\ Phys.\ B {\bf 444}, 125 (1995).
[hep-th/9503179].
}

\lref\PasquettiFJ{
  S.~Pasquetti,
  ``Factorisation of N = 2 Theories on the Squashed 3-Sphere,''
JHEP {\bf 1204}, 120 (2012).
[arXiv:1111.6905 [hep-th]].
}

\lref\BeemMB{
  C.~Beem, T.~Dimofte and S.~Pasquetti,
  ``Holomorphic Blocks in Three Dimensions,''
[arXiv:1211.1986 [hep-th]].
}

\lref\SeibergPQ{
  N.~Seiberg,
  ``Electric - magnetic duality in supersymmetric nonAbelian gauge theories,''
Nucl.\ Phys.\ B {\bf 435}, 129 (1995).
[hep-th/9411149].
}

\lref\AharonyBX{
  O.~Aharony, A.~Hanany, K.~A.~Intriligator, N.~Seiberg and M.~J.~Strassler,
  ``Aspects of N=2 supersymmetric gauge theories in three-dimensions,''
Nucl.\ Phys.\ B {\bf 499}, 67 (1997).
[hep-th/9703110].
}

\lref\IntriligatorNE{
  K.~A.~Intriligator and P.~Pouliot,
  ``Exact superpotentials, quantum vacua and duality in supersymmetric SP(N(c)) gauge theories,''
Phys.\ Lett.\ B {\bf 353}, 471 (1995).
[hep-th/9505006].
}

\lref\GaiottoKFA{
  D.~Gaiotto, A.~Kapustin, N.~Seiberg and B.~Willett,
  ``Generalized Global Symmetries,''
JHEP {\bf 1502}, 172 (2015).
[arXiv:1412.5148 [hep-th]].
}

\lref\KarchUX{
  A.~Karch,
  ``Seiberg duality in three-dimensions,''
Phys.\ Lett.\ B {\bf 405}, 79 (1997).
[hep-th/9703172].
}

\lref\SafdiRE{
  B.~R.~Safdi, I.~R.~Klebanov and J.~Lee,
  ``A Crack in the Conformal Window,''
[arXiv:1212.4502 [hep-th]].
}

\lref\SchweigertTG{
  C.~Schweigert,
  ``On moduli spaces of flat connections with nonsimply connected structure group,''
Nucl.\ Phys.\ B {\bf 492}, 743 (1997).
[hep-th/9611092].
}

\lref\GiveonZN{
  A.~Giveon and D.~Kutasov,
  ``Seiberg Duality in Chern--Simons Theory,''
Nucl.\ Phys.\ B {\bf 812}, 1 (2009).
[arXiv:0808.0360 [hep-th]].
}

\lref\GaiottoBE{
  D.~Gaiotto, G.~W.~Moore and A.~Neitzke,
  ``Framed BPS States,''
[arXiv:1006.0146 [hep-th]].
}

\lref\AldayRS{
  L.~F.~Alday, M.~Bullimore and M.~Fluder,
  ``On S-duality of the Superconformal Index on Lens Spaces and 2d TQFT,''
JHEP {\bf 1305}, 122 (2013).
[arXiv:1301.7486 [hep-th]].
}

\lref\RazamatJXA{
  S.~S.~Razamat and M.~Yamazaki,
  ``S-duality and the N=2 Lens Space Index,''
[arXiv:1306.1543 [hep-th]].
}

\lref\NiarchosAH{
  V.~Niarchos,
  ``Seiberg dualities and the 3d/4d connection,''
JHEP {\bf 1207}, 075 (2012).
[arXiv:1205.2086 [hep-th]].
}

\lref\almost{
  A.~Borel, R.~Friedman, J.~W.~Morgan,
  ``Almost commuting elements in compact Lie groups,''
arXiv:math/9907007.
}

\lref\AharonyDHA{
  O.~Aharony, S.~S.~Razamat, N.~Seiberg and B.~Willett,
  ``$3d$ dualities from $4d$ dualities,''
JHEP {\bf 1307}, 149 (2013).
[arXiv:1305.3924 [hep-th]].
}

\lref\KapustinJM{
  A.~Kapustin and B.~Willett,
  ``Generalized Superconformal Index for Three Dimensional Field Theories,''
[arXiv:1106.2484 [hep-th]].
}

\lref\AharonyGP{
  O.~Aharony,
  ``IR duality in d = 3 N=2 supersymmetric USp(2N(c)) and U(N(c)) gauge theories,''
Phys.\ Lett.\ B {\bf 404}, 71 (1997).
[hep-th/9703215].
}

\lref\ParkWTA{
  J.~Park and K.~J.~Park,
  ``Seiberg-like Dualities for 3d N=2 Theories with SU(N) gauge group,''
JHEP {\bf 1310}, 198 (2013).
[arXiv:1305.6280 [hep-th]].
}

\lref\FestucciaWS{
  G.~Festuccia and N.~Seiberg,
  ``Rigid Supersymmetric Theories in Curved Superspace,''
JHEP {\bf 1106}, 114 (2011).
[arXiv:1105.0689 [hep-th]].
}

\lref\RomelsbergerEG{
  C.~Romelsberger,
  ``Counting chiral primaries in N = 1, d=4 superconformal field theories,''
Nucl.\ Phys.\ B {\bf 747}, 329 (2006).
[hep-th/0510060].
}

\lref\KapustinKZ{
  A.~Kapustin, B.~Willett and I.~Yaakov,
  ``Exact Results for Wilson Loops in Superconformal Chern--Simons Theories with Matter,''
JHEP {\bf 1003}, 089 (2010).
[arXiv:0909.4559 [hep-th]].
}

\lref\DolanQI{
  F.~A.~Dolan and H.~Osborn,
  ``Applications of the Superconformal Index for Protected Operators and q-Hypergeometric Identities to N=1 Dual Theories,''
Nucl.\ Phys.\ B {\bf 818}, 137 (2009).
[arXiv:0801.4947 [hep-th]].
}

\lref\GaddeIA{
  A.~Gadde and W.~Yan,
  ``Reducing the 4d Index to the $S^3$ Partition Function,''
JHEP {\bf 1212}, 003 (2012).
[arXiv:1104.2592 [hep-th]].
}

\lref\DolanRP{
  F.~A.~H.~Dolan, V.~P.~Spiridonov and G.~S.~Vartanov,
  ``From 4d superconformal indices to 3d partition functions,''
Phys.\ Lett.\ B {\bf 704}, 234 (2011).
[arXiv:1104.1787 [hep-th]].
}

\lref\ImamuraUW{
  Y.~Imamura,
 ``Relation between the 4d superconformal index and the $S^3$ partition function,''
JHEP {\bf 1109}, 133 (2011).
[arXiv:1104.4482 [hep-th]].
}

\lref\LeighEP{
  R.~G.~Leigh and M.~J.~Strassler,
  ``Exactly marginal operators and duality in four-dimensional N=1 supersymmetric gauge theory,''
Nucl.\ Phys.\ B {\bf 447}, 95 (1995).
[hep-th/9503121].
}

\lref\HamaEA{
  N.~Hama, K.~Hosomichi and S.~Lee,
  ``SUSY Gauge Theories on Squashed Three-Spheres,''
JHEP {\bf 1105}, 014 (2011).
[arXiv:1102.4716 [hep-th]].
}

\lref\GaddeEN{
  A.~Gadde, L.~Rastelli, S.~S.~Razamat and W.~Yan,
  ``On the Superconformal Index of N=1 IR Fixed Points: A Holographic Check,''
JHEP {\bf 1103}, 041 (2011).
[arXiv:1011.5278 [hep-th]].
}

\lref\EagerHX{
  R.~Eager, J.~Schmude and Y.~Tachikawa,
  ``Superconformal Indices, Sasaki-Einstein Manifolds, and Cyclic Homologies,''
[arXiv:1207.0573 [hep-th]].
}

\lref\AffleckAS{
  I.~Affleck, J.~A.~Harvey and E.~Witten,
  ``Instantons and (Super)Symmetry Breaking in (2+1)-Dimensions,''
Nucl.\ Phys.\ B {\bf 206}, 413 (1982)..
}

\lref\SeibergPQ{
  N.~Seiberg,
  ``Electric - magnetic duality in supersymmetric nonAbelian gauge theories,''
Nucl.\ Phys.\ B {\bf 435}, 129 (1995).
[hep-th/9411149].
}

\lref\debult{
  F.~van~de~Bult,
  ``Hyperbolic Hypergeometric Functions,''
University of Amsterdam Ph.D. thesis.
}

\lref\Shamirthesis{
  I.~Shamir,
  ``Aspects of three dimensional Seiberg duality,''
  M. Sc. thesis submitted to the Weizmann Institute of Science, April 2010.
  }

\lref\slthreeZ{
  J.~Felder, A.~Varchenko,
  ``The elliptic gamma function and $SL(3,Z) \times Z^3$,'' $\;\;$
[arXiv:math/0001184].
}

\lref\BeniniNC{
  F.~Benini, T.~Nishioka and M.~Yamazaki,
  ``4d Index to 3d Index and 2d TQFT,''
Phys.\ Rev.\ D {\bf 86}, 065015 (2012).
[arXiv:1109.0283 [hep-th]].
}

\lref\GaiottoWE{
  D.~Gaiotto,
  ``N=2 dualities,''
  JHEP {\bf 1208}, 034 (2012).
  [arXiv:0904.2715 [hep-th]].
}

\lref\SpiridonovZA{
  V.~P.~Spiridonov and G.~S.~Vartanov,
  ``Elliptic Hypergeometry of Supersymmetric Dualities,''
Commun.\ Math.\ Phys.\  {\bf 304}, 797 (2011).
[arXiv:0910.5944 [hep-th]].
}

\lref\BeniniMF{
  F.~Benini, C.~Closset and S.~Cremonesi,
  ``Comments on 3d Seiberg-like dualities,''
JHEP {\bf 1110}, 075 (2011).
[arXiv:1108.5373 [hep-th]].
}

\lref\ClossetVP{
  C.~Closset, T.~T.~Dumitrescu, G.~Festuccia, Z.~Komargodski and N.~Seiberg,
  ``Comments on Chern--Simons Contact Terms in Three Dimensions,''
JHEP {\bf 1209}, 091 (2012).
[arXiv:1206.5218 [hep-th]].
}

\lref\SpiridonovHF{
  V.~P.~Spiridonov and G.~S.~Vartanov,
  ``Elliptic hypergeometry of supersymmetric dualities II. Orthogonal groups, knots, and vortices,''
[arXiv:1107.5788 [hep-th]].
}

\lref\SpiridonovWW{
  V.~P.~Spiridonov and G.~S.~Vartanov,
  ``Elliptic hypergeometric integrals and 't Hooft anomaly matching conditions,''
JHEP {\bf 1206}, 016 (2012).
[arXiv:1203.5677 [hep-th]].
}

\lref\RazamatGRO{
  S.~S.~Razamat and G.~Zafrir,
  ``Compactification of 6d minimal SCFTs on Riemann surfaces,''
[arXiv:1806.09196 [hep-th]].
}

\lref\DimoftePY{
  T.~Dimofte, D.~Gaiotto and S.~Gukov,
Adv.\ Theor.\ Math.\ Phys.\  {\bf 17}, no. 5, 975 (2013).
[arXiv:1112.5179 [hep-th]].
}

\lref\KimWB{
  S.~Kim,
  ``The Complete superconformal index for N=6 Chern--Simons theory,''
Nucl.\ Phys.\ B {\bf 821}, 241 (2009), [Erratum-ibid.\ B {\bf 864}, 884 (2012)].
[arXiv:0903.4172 [hep-th]].
}

\lref\WillettGP{
  B.~Willett and I.~Yaakov,
  ``N=2 Dualities and Z Extremization in Three Dimensions,''
[arXiv:1104.0487 [hep-th]].
}

\lref\ImamuraSU{
  Y.~Imamura and S.~Yokoyama,
  ``Index for three dimensional superconformal field theories with general R-charge assignments,''
JHEP {\bf 1104}, 007 (2011).
[arXiv:1101.0557 [hep-th]].
}

\lref\FreedYA{
  D.~S.~Freed, G.~W.~Moore and G.~Segal,
  ``The Uncertainty of Fluxes,''
Commun.\ Math.\ Phys.\  {\bf 271}, 247 (2007).
[hep-th/0605198].
}

\lref\HwangQT{
  C.~Hwang, H.~Kim, K.~-J.~Park and J.~Park,
  ``Index computation for 3d Chern--Simons matter theory: test of Seiberg-like duality,''
JHEP {\bf 1109}, 037 (2011).
[arXiv:1107.4942 [hep-th]].
}

\lref\GreenDA{
  D.~Green, Z.~Komargodski, N.~Seiberg, Y.~Tachikawa and B.~Wecht,
  ``Exactly Marginal Deformations and Global Symmetries,''
JHEP {\bf 1006}, 106 (2010).
[arXiv:1005.3546 [hep-th]].
}

\lref\GaiottoXA{
  D.~Gaiotto, L.~Rastelli and S.~S.~Razamat,
  ``Bootstrapping the superconformal index with surface defects,''
[arXiv:1207.3577 [hep-th]].
}

\lref\IntriligatorID{
  K.~A.~Intriligator and N.~Seiberg,
  ``Duality, monopoles, dyons, confinement and oblique confinement in supersymmetric SO(N(c)) gauge theories,''
Nucl.\ Phys.\ B {\bf 444}, 125 (1995).
[hep-th/9503179].
}

\lref\SeibergNZ{
  N.~Seiberg and E.~Witten,
  ``Gauge dynamics and compactification to three-dimensions,''
In *Saclay 1996, The mathematical beauty of physics* 333-366.
[hep-th/9607163].
}

\lref\KinneyEJ{
  J.~Kinney, J.~M.~Maldacena, S.~Minwalla and S.~Raju,
  ``An Index for 4 dimensional super conformal theories,''
  Commun.\ Math.\ Phys.\  {\bf 275}, 209 (2007).
  [hep-th/0510251].
}

\lref\NakayamaUR{
  Y.~Nakayama,
  ``Index for supergravity on AdS(5) x T**1,1 and conifold gauge theory,''
Nucl.\ Phys.\ B {\bf 755}, 295 (2006).
[hep-th/0602284].
}

\lref\GaddeKB{
  A.~Gadde, E.~Pomoni, L.~Rastelli and S.~S.~Razamat,
  ``S-duality and 2d Topological QFT,''
JHEP {\bf 1003}, 032 (2010).
[arXiv:0910.2225 [hep-th]].
}

\lref\GaddeTE{
  A.~Gadde, L.~Rastelli, S.~S.~Razamat and W.~Yan,
  ``The Superconformal Index of the $E_6$ SCFT,''
JHEP {\bf 1008}, 107 (2010).
[arXiv:1003.4244 [hep-th]].
}

\lref\AharonyCI{
  O.~Aharony and I.~Shamir,
  ``On $O(N_c)$ d=3 N=2 supersymmetric QCD Theories,''
JHEP {\bf 1112}, 043 (2011).
[arXiv:1109.5081 [hep-th]].
}

\lref\GiveonSR{
  A.~Giveon and D.~Kutasov,
  ``Brane dynamics and gauge theory,''
Rev.\ Mod.\ Phys.\  {\bf 71}, 983 (1999).
[hep-th/9802067].
}

\lref\BeniniNC{
  F.~Benini, T.~Nishioka and M.~Yamazaki,
  ``4d Index to 3d Index and 2d TQFT,''
Phys.\ Rev.\ D {\bf 86}, 065015 (2012).
[arXiv:1109.0283 [hep-th]].
}

\lref\SpiridonovQV{
  V.~P.~Spiridonov and G.~S.~Vartanov,
  ``Superconformal indices of ${\cal N}=4$ SYM field theories,''
Lett.\ Math.\ Phys.\  {\bf 100}, 97 (2012).
[arXiv:1005.4196 [hep-th]].
}
\lref\GaddeUV{
  A.~Gadde, L.~Rastelli, S.~S.~Razamat and W.~Yan,
  ``Gauge Theories and Macdonald Polynomials,''
Commun.\ Math.\ Phys.\  {\bf 319}, 147 (2013).
[arXiv:1110.3740 [hep-th]].
}
\lref\KapustinGH{
  A.~Kapustin,
  ``Seiberg-like duality in three dimensions for orthogonal gauge groups,''
[arXiv:1104.0466 [hep-th]].
}

\lref\AharonyKMA{
  O.~Aharony, S.~S.~Razamat, N.~Seiberg and B.~Willett,
  ``$3d$ dualities from $4d$ dualities for orthogonal groups,''
JHEP {\bf 1308}, 099 (2013).
[arXiv:1307.0511 [hep-th]].
}

\lref\AharonyHDA{
  O.~Aharony, N.~Seiberg and Y.~Tachikawa,
  ``Reading between the lines of four-dimensional gauge theories,''
JHEP {\bf 1308}, 115 (2013).
[arXiv:1305.0318 [hep-th]].
}

\lref\amacassi{
A. Amariti and L. Cassia,
``$USp(2N_c)$ SQCD$_3$ with antisymmetric: dualities and symmetry enhancements,''
[arXiv:1809.03796 [hep-th]].
}

\lref\benve{
S. Benvenuti,
``A tale of exceptional 3d dualities,''
[arXiv:1809.03925 [hep-th]].
}

\lref\WittenNV{
  E.~Witten,
  ``Supersymmetric index in four-dimensional gauge theories,''
Adv.\ Theor.\ Math.\ Phys.\  {\bf 5}, 841 (2002).
[hep-th/0006010].
}

\lref\GaddeUV{
  A.~Gadde, L.~Rastelli, S.~S.~Razamat and W.~Yan,
  ``Gauge Theories and Macdonald Polynomials,''
Commun.\ Math.\ Phys.\  {\bf 319}, 147 (2013).
[arXiv:1110.3740 [hep-th]].
}

\lref\GaddeIK{
  A.~Gadde, L.~Rastelli, S.~S.~Razamat and W.~Yan,
  ``The 4d Superconformal Index from q-deformed 2d Yang-Mills,''
Phys.\ Rev.\ Lett.\  {\bf 106}, 241602 (2011).
[arXiv:1104.3850 [hep-th]].
}

\lref\GaiottoXA{
  D.~Gaiotto, L.~Rastelli and S.~S.~Razamat,
  ``Bootstrapping the superconformal index with surface defects,''
JHEP {\bf 1301}, 022 (2013).
[arXiv:1207.3577 [hep-th]].
}

\lref\GaiottoUQ{
  D.~Gaiotto and S.~S.~Razamat,
  ``Exceptional Indices,''
JHEP {\bf 1205}, 145 (2012).
[arXiv:1203.5517 [hep-th]].
}

\lref\RazamatUV{
  S.~S.~Razamat,
  ``On a modular property of N=2 superconformal theories in four dimensions,''
JHEP {\bf 1210}, 191 (2012).
[arXiv:1208.5056 [hep-th]].
}

\lref\noumi{
  Y.~Komori, M.~Noumi, J.~Shiraishi,
  ``Kernel Functions for Difference Operators of Ruijsenaars Type and Their Applications,''
SIGMA 5 (2009), 054.
[arXiv:0812.0279 [math.QA]].
}

\lref\RazamatJXA{
  S.~S.~Razamat and M.~Yamazaki,
  ``S-duality and the N=2 Lens Space Index,''
[arXiv:1306.1543 [hep-th]].
}

\lref\RazamatOPA{
  S.~S.~Razamat and B.~Willett,
  ``Global Properties of Supersymmetric Theories and the Lens Space,''
Commun.\ Math.\ Phys.\  {\bf 334}, no. 2, 661 (2015).
[arXiv:1307.4381 [hep-th]].
}

\lref\GaddeTE{
  A.~Gadde, L.~Rastelli, S.~S.~Razamat and W.~Yan,
  ``The Superconformal Index of the $E_6$ SCFT,''
JHEP {\bf 1008}, 107 (2010).
[arXiv:1003.4244 [hep-th]].
}

\lref\deBult{
  F.~J.~van~de~Bult,
  ``An elliptic hypergeometric integral with $W(F_4)$ symmetry,''
The Ramanujan Journal, Volume 25, Issue 1 (2011).
[arXiv:0909.4793[math.CA]].
}

\lref\GaddeKB{
  A.~Gadde, E.~Pomoni, L.~Rastelli and S.~S.~Razamat,
  ``S-duality and 2d Topological QFT,''
JHEP {\bf 1003}, 032 (2010).
[arXiv:0910.2225 [hep-th]].
}

\lref\ArgyresCN{
  P.~C.~Argyres and N.~Seiberg,
  ``S-duality in N=2 supersymmetric gauge theories,''
JHEP {\bf 0712}, 088 (2007).
[arXiv:0711.0054 [hep-th]].
}

\lref\RazamatPTA{
  S.~S.~Razamat and B.~Willett,
  ``Down the rabbit hole with theories of class $ {\cal S} $,''
JHEP {\bf 1410}, 99 (2014).
[arXiv:1403.6107 [hep-th]].
}

\lref\SpirWarnaar{
  V.~P.~Spiridonov and S.~O.~Warnaar,
  ``Inversions of integral operators and elliptic beta integrals on root systems,''
Adv. Math. 207 (2006), 91-132
[arXiv:math/0411044].
}

\lref\GaiottoHG{
  D.~Gaiotto, G.~W.~Moore and A.~Neitzke,
  ``Wall-crossing, Hitchin Systems, and the WKB Approximation,''
[arXiv:0907.3987 [hep-th]].
}

\lref\ClossetZGF{
  C.~Closset, H.~Kim and B.~Willett,
  ``Supersymmetric partition functions and the three-dimensional A-twist,''
JHEP {\bf 1703}, 074 (2017).
[arXiv:1701.03171 [hep-th]].
}

\lref\RuijsenaarsVQ{
  S.~N.~M.~Ruijsenaars and H.~Schneider,
  ``A New Class Of Integrable Systems And Its Relation To Solitons,''
Annals Phys.\  {\bf 170}, 370 (1986).
}

\lref\BaggioMAS{
  M.~Baggio, N.~Bobev, S.~M.~Chester, E.~Lauria and S.~S.~Pufu,
  ``Decoding a Three-Dimensional Conformal Manifold,''
JHEP {\bf 1802}, 062 (2018).
[arXiv:1712.02698 [hep-th]].
}

\lref\AharonyGP{
  O.~Aharony,
  ``IR duality in $d = 3$ $\cal{N} =$ 2 supersymmetric $USp(2N_c)$ and $U(N_c)$ gauge theories,''
Phys.\ Lett.\ B {\bf 404}, 71 (1997).
[hep-th/9703215].
}

\lref\RuijsenaarsPP{
  S.~N.~M.~Ruijsenaars,
  ``Complete Integrability Of Relativistic Calogero-moser Systems And Elliptic Function Identities,''
Commun.\ Math.\ Phys.\  {\bf 110}, 191 (1987).
}

\lref\HallnasNB{
  M.~Hallnas and S.~Ruijsenaars,
  ``Kernel functions and Baecklund transformations for relativistic Calogero-Moser and Toda systems,''
J.\ Math.\ Phys.\  {\bf 53}, 123512 (2012).
}

\lref\kernelA{
S.~Ruijsenaars,
  ``Elliptic integrable systems of Calogero-Moser type: Some new results on joint eigenfunctions'', in Proceedings of the 2004 Kyoto Workshop on "Elliptic integrable systems", (M. Noumi, K. Takasaki, Eds.), Rokko Lectures in Math., no. 18, Dept. of Math., Kobe Univ.
}

\lref\ellRSreview{
Y.~Komori and S.~Ruijsenaars,
  ``Elliptic integrable systems of Calogero-Moser type: A survey'', in Proceedings of the 2004 Kyoto Workshop on "Elliptic integrable systems", (M. Noumi, K. Takasaki, Eds.), Rokko Lectures in Math., no. 18, Dept. of Math., Kobe Univ.
}

\lref\AharonyMJS{
  O.~Aharony,
  ``Baryons, monopoles and dualities in Chern--Simons-matter theories,''
JHEP {\bf 1602}, 093 (2016).
[arXiv:1512.00161 [hep-th]].
}

\lref\langmann{
E.~Langmann,
  ``An explicit solution of the (quantum) elliptic Calogero-Sutherland model'', [arXiv:math-ph/0407050].
}

\lref\JafferisUN{
  D.~L.~Jafferis,
  ``The Exact Superconformal R-Symmetry Extremizes Z,''
JHEP {\bf 1205}, 159 (2012).
[arXiv:1012.3210 [hep-th]].
}

\lref\TachikawaWI{
  Y.~Tachikawa,
  ``$4d$ partition function on $S^1 \times S^3$ and $2d$ Yang--Mills with nonzero area,''
PTEP {\bf 2013}, 013B01 (2013).
[arXiv:1207.3497 [hep-th]].
}

\lref\WillettADV{
  B.~Willett,
  ``Localization on three-dimensional manifolds,''
J.\ Phys.\ A {\bf 50}, no. 44, 443006 (2017).
[arXiv:1608.02958 [hep-th]].
}

\lref\MinahanFG{
  J.~A.~Minahan and D.~Nemeschansky,
  ``An $\cal{N} =$ 2 superconformal fixed point with $E_6$ global symmetry,''
Nucl.\ Phys.\ B {\bf 482}, 142 (1996).
[hep-th/9608047].
}

\lref\BeniniDUD{
  F.~Benini, S.~Benvenuti and S.~Pasquetti,
  ``SUSY monopole potentials in 2+1 dimensions,''
JHEP {\bf 1708}, 086 (2017).
[arXiv:1703.08460 [hep-th]].
}

\lref\BenvenutiWET{
S.~Benvenuti and S.~Pasquetti,
``3d $ \cal{N}=$ 2 mirror symmetry, pq-webs and monopole superpotentials,''
JHEP {\bf 1608}, 136 (2016).
[arXiv:1605.02675[hep-th]].
}

\lref\AmaritiGDC{
A.~Amariti, I.~Garozzo and N.~Mekareeya,
``New $3d$ $\cal{N}=$ 2 Dualities from Quadratic Monopoles,''
[arXiv:1806.01356[hep-th]].
}

\lref\AldayKDA{
  L.~F.~Alday, M.~Bullimore, M.~Fluder and L.~Hollands,
  ``Surface defects, the superconformal index and q-deformed Yang-Mills,''
[arXiv:1303.4460 [hep-th]].
}

\lref\PestunZXK{
  V.~Pestun {\it et al.},
  ``Localization techniques in quantum field theories,''
J.\ Phys.\ A {\bf 50}, no. 44, 440301 (2017).
[arXiv:1608.02952 [hep-th]].
}

\lref\FukudaJR{
  Y.~Fukuda, T.~Kawano and N.~Matsumiya,
  ``5D SYM and 2D q-Deformed YM,''
Nucl.\ Phys.\ B {\bf 869}, 493 (2013).
[arXiv:1210.2855 [hep-th]].
}

\lref\ChoiOHN{
  C.~Choi, M.~Rocek and A.~Sharon,
  ``Dualities and Phases of 3D $\cal{N}=$ 1 SQCD,''
[arXiv:1808.02184 [hep-th]].
}

\lref\XieHS{
  D.~Xie,
  ``General Argyres-Douglas Theory,''
JHEP {\bf 1301}, 100 (2013).
[arXiv:1204.2270 [hep-th]].
}

\lref\KapustinSim{
A.~Kapustin,  2011 Simons Summer Workshop seminar. A video of this talk can be found at
{\tt 
http://scgp.stonybrook.edu/video$\_$portal/video.php?id=780  
}.

}

\lref\DrukkerSR{
  N.~Drukker, T.~Okuda and F.~Passerini,
  ``Exact results for vortex loop operators in 3d supersymmetric theories,''
[arXiv:1211.3409 [hep-th]].
}

\lref\qinteg{
  M.~Rahman, A.~Verma,
  ``A q-integral representation of Rogers' q-ultraspherical polynomials and some applications,''
Constructive Approximation
1986, Volume 2, Issue 1.
}

\lref\AharonyUYA{
  O.~Aharony and D.~Fleischer,
  ``IR Dualities in General 3d Supersymmetric SU(N) QCD Theories,''
JHEP {\bf 1502}, 162 (2015).
[arXiv:1411.5475 [hep-th]].
}

\lref\qintegOK{
  A.~Okounkov,
  ``(Shifted) Macdonald Polynomials: q-Integral Representation and Combinatorial Formula,''
Compositio Mathematica
June 1998, Volume 112, Issue 2. 
[arXiv:q-alg/9605013].
}

\lref\RazamatGZX{
  S.~S.~Razamat and G.~Zafrir,
  ``Exceptionally simple exceptional models,''
JHEP {\bf 1611}, 061 (2016).
[arXiv:1609.02089 [hep-th]].
}

\lref\macNest{
 H.~Awata, S.~Odake, J.~Shiraishi,
  ``Integral Representations of the Macdonald Symmetric Functions,''
Commun. Math. Phys. 179 (1996) 647.
[arXiv:q-alg/9506006].
}

\lref\BeemYN{
  C.~Beem and A.~Gadde,
  ``The $N=1$ superconformal index for class $S$ fixed points,''
JHEP {\bf 1404}, 036 (2014).
[arXiv:1212.1467 [hep-th]].
}

\lref\BenvenutiGA{
  S.~Benvenuti and S.~Pasquetti,
  ``3D-partition functions on the sphere: exact evaluation and mirror symmetry,''
JHEP {\bf 1205}, 099 (2012).
[arXiv:1105.2551 [hep-th]].
}

\lref\GaiottoYJH{
  D.~Gaiotto, Z.~Komargodski and J.~Wu,
  ``Curious Aspects of Three-Dimensional ${\cal N}=1$ SCFTs,''
[arXiv:1804.02018 [hep-th]].
}

\lref\GaddeFMA{
  A.~Gadde, K.~Maruyoshi, Y.~Tachikawa and W.~Yan,
  ``New N=1 Dualities,''
JHEP {\bf 1306}, 056 (2013).
[arXiv:1303.0836 [hep-th]].
}

\lref\BashmakovWTS{
  V.~Bashmakov, J.~Gomis, Z.~Komargodski and A.~Sharon,
  ``Phases of $ {\cal N}=1 $ theories in 2 + 1 dimensions,''
JHEP {\bf 1807}, 123 (2018).
[arXiv:1802.10130 [hep-th]].
}

\lref\GaiottoAK{
  D.~Gaiotto and E.~Witten,
  ``S-Duality of Boundary Conditions In N=4 Super Yang-Mills Theory,''
Adv.\ Theor.\ Math.\ Phys.\  {\bf 13}, no. 3, 721 (2009).
[arXiv:0807.3720 [hep-th]].
}

\lref\GangWEK{
  D.~Gang and K.~Yonekura,
  ``Symmetry enhancement and closing of knots in 3d/3d correspondence,''
JHEP {\bf 1807}, 145 (2018).
[arXiv:1803.04009 [hep-th]].
}

\lref\GorskyTN{
  A.~Gorsky,
  ``Dualities in integrable systems and N=2 SUSY theories,''
J.\ Phys.\ A {\bf 34}, 2389 (2001).
[hep-th/9911037].
}

\lref\GangLSR{
  D.~Gang, Y.~Tachikawa and K.~Yonekura,
  ``Smallest 3d hyperbolic manifolds via simple 3d theories,''
Phys.\ Rev.\ D {\bf 96}, no. 6, 061701 (2017).
[arXiv:1706.06292 [hep-th]].
}

\lref\FockAE{
  V.~Fock, A.~Gorsky, N.~Nekrasov and V.~Rubtsov,
  ``Duality in integrable systems and gauge theories,''
JHEP {\bf 0007}, 028 (2000).
[hep-th/9906235].
}

\lref\RazamatOPA{
  S.~S.~Razamat and B.~Willett,
  ``Global Properties of Supersymmetric Theories and the Lens Space,''
Commun.\ Math.\ Phys.\  {\bf 334}, no. 2, 661 (2015).
[arXiv:1307.4381 [hep-th]].
}

\lref\BeniniUMH{
  F.~Benini and S.~Benvenuti,
  ``$\cal{N}=$ 1 QED in 2+1 dimensions,''
[arXiv:1803.01784 [hep-th]].
} 
 
\lref\BeniniBHK{
  F.~Benini and S.~Benvenuti,
  ``$\cal{N}=$ 1 QED in 2+1 dimensions: Dualities and enhanced symmetries,''
[arXiv:1804.05707 [hep-th]].
}

\lref\SeibergGMD{
  N.~Seiberg, T.~Senthil, C.~Wang and E.~Witten,
  ``A Duality Web in 2+1 Dimensions and Condensed Matter Physics,''
Annals Phys.\  {\bf 374}, 395 (2016).
[arXiv:1606.01989 [hep-th]].
}

\lref\CsakiCU{
  C.~Csaki, M.~Schmaltz, W.~Skiba and J.~Terning,
  ``Selfdual N=1 SUSY gauge theories,''
Phys.\ Rev.\ D {\bf 56}, 1228 (1997).
[hep-th/9701191].
}

\lref\BeniniUI{
  F.~Benini and S.~Cremonesi,
[arXiv:1206.2356 [hep-th]].
}

\lref\GiacomelliVGK{
S.~Giacomelli and N.~Mekareeya,
``Mirror theories of $3d$ $\cal{N}  =$ 2 SQCD,''
JHEP {\bf 1803}, 123 (2018).
[hep-th/1711.11525].
}

\lref\DoroudXW{
  N.~Doroud, J.~Gomis, B.~Le Floch and S.~Lee,
JHEP {\bf 1305}, 093 (2013).
[arXiv:1206.2606 [hep-th]].
}

\lref\GomisWY{
  J.~Gomis and S.~Lee,
JHEP {\bf 1304}, 019 (2013).
[arXiv:1210.6022 [hep-th]].
}

\lref\AganagicUW{
  M.~Aganagic, K.~Hori, A.~Karch and D.~Tong,
JHEP {\bf 0107}, 022 (2001).
[hep-th/0105075].
}

\lref\HoriKT{
  K.~Hori and C.~Vafa,
[hep-th/0002222].
}

\Title{\vbox{\baselineskip12pt
}}
{\vbox{
\centerline{Chiral $3d$ $SU(3)$ SQCD and ${\cal N}=2$ mirror duality}
}
}
\centerline{Marco Fazzi,$^{a,b}$ Assaf Lanir,$^a$ Shlomo S. Razamat,$^a$ and Orr Sela$^a$}
\bigskip
\centerline{{\it $^a$Department of Physics, Technion, Haifa, 32000, Israel}}
\centerline{{\it $^b$Department of Mathematics and Haifa Research Center for Theoretical Physics}}
\centerline{{\it and Astrophysics, University of Haifa, Haifa, 31905, Israel}}

\vskip.1in \vskip.2in \centerline{\bf Abstract}

Recently a very interesting three-dimensional  ${\cal N}=2$ supersymmetric theory with $SU(3)$ global symmetry was discussed by several authors.  We  denote this model by $T_x$.
This  was conjectured to have two dual descriptions, one with explicit supersymmetry and emergent  flavor symmetry and the other with explicit flavor symmetry and emergent supersymmetry. We discuss a third description of the model which has both flavor symmetry and supersymmetry manifest. We then investigate models which can be constructed by using $T_x$ as a building block gauging the global symmetry and  paying special attention to the global structure of the gauge group. We conjecture several  cases of ${\cal N}=2$ mirror dualities involving such constructions with the dual being either a  simple ${\cal N}=2$ Wess--Zumino model or a discrete  gauging thereof.

\vskip.2in

\noindent

\vfill

\Date{August 2018}

\newsec{Introduction and Discussion}

Three dimensions is quantum field theorist's paradise. On the one hand it is easy to build simple asymptotically free field theories which flow to interacting fixed points,  and on the other we have a lot of control over many such models. This is to be contrasted with the situation in higher dimensions, where the number of interesting models in the IR with simple UV Lagrangians decreases, and in lower dimensions, where many subtle effects pertaining to the vacuum structure appear. Moreover, in principle, models in three dimensions can be engineered as effective descriptions of real-world condensed matter systems in a lab.

The situation is particularly beneficial with supersymmetric theories, where we have a plethora of exact computations we can perform for ${\cal N}=2$ supersymmetric cases (see e.g. \KapustinKZ\WillettADV\ClossetZGF). Recently even models with ${\cal N}=1$ supersymmetry have led to exact results \BashmakovWTS\BeniniUMH\GaiottoYJH\BeniniBHK\ChoiOHN. Moreover, the progress with understanding the supersymmetric models and also the large-$N$ models, as summarized in \AharonyMJS, has led to a remarkable progress in understanding non supersymmetric Chern--Simons theories, see e.g. \SeibergGMD.

In this note we stay in the ${\cal N}=2$ supersymmetric domain and discuss some  effects motivated by recent progress with less supersymmetric theories. In particular, a very interesting model, which modulo contact terms we will call $T_x$, has been conjectured to have two descriptions \GaiottoYJH\BeniniBHK. One with manifest $SU(3)$ global symmetry and ${\cal N}=1$ supersymmetry, which enhances to ${\cal N}=2$ in the IR. The other as an ${\cal N}=2$ theory with $U(1)\times SU(2)$ global symmetry enhancing to $SU(3)$ in the IR.
The fact that the symmetry enhances to $SU(3)$ was also obtained from geometric considerations in the context of the $3d/3d$ correspondence \DimoftePY\ in \GangLSR\GangWEK. We will suggest here a third description (from which yet another can be derived using a by now well-known IR duality \BeniniMF\AharonyUYA) which has both ${\cal N}=2$ supersymmetry and $SU(3)$ global symmetry in the UV. The description is an $SU(3)$ Chern--Simons model with level $5/2$ and a single chiral field in the bi-fundamental representation of the gauge and flavor $SU(3)$, supplemented by a baryon superpotential. We will give evidence for the duality by comparing the superconformal index and three-sphere partition function of this model and of the description with non manifest flavor symmetry but manifest supersymmetry.

In the second part of the note we will  construct theories which do not have any continuous symmetries by gauging with Chern--Simons terms the diagonal global symmetry of several $T_x$ models. As the model $T_x$ has only matter charged in representations with $N$-ality zero under $SU(3)$, both gauging $SU(3)$ and $SU(3)/\Z_3$ is possible. In three dimensions the latter possibility leads to models with $\Z_3$ zero-form global symmetry. This is to be contrasted with four dimensions where such a choice of global structure affects the spectrum of line operators as it affects one-form symmetries. See \GaiottoKFA\ 
and \AharonyHDA\ 
for recent discussions. The operators which are charged under such discrete symmetries in three dimensions are gauge-invariant monopoles.
 We will mainly focus on the case with $SU(3)/\Z_3$ gauging and construct theories which have  discrete global symmetry. For several examples of lowest possible values of Chern--Simons levels and low number of copies we conjecture that such theories are dual to Wess--Zumino models with ${\cal N}=2$ supersymmetry and cubic superpotential interactions. Note that these interactions preserve a $\Z_3$ symmetry. Performing the gauging with $SU(3)$ will result in such duals with the $\Z_3$ symmetry gauged. The main evidence we give for the conjectured dualities is again by comparing the supersymmetric indices.

We observe several other interesting features from our results. For instance, taking four copies of $T_x$ and gauging $SU(3)$ with level two we obtain evidence that the model is dual to $T_x$ with the $SU(3)$ symmetry emerging in IR. Another observation is that the basic monopole operators in many examples we study here are counted by Catalan numbers, and it would be interesting to understand whether mirror models with such property, that is Catalan numbers counting operators built from fundamental fields, can be considered. Finally, $SU(3)$ plays a special role in our construction. This is mainly because it is easy to construct $SU(3)$-invariant relevant superpotential with matter in fundamental representations. 
 The group $SU(3)$ plays a special role in four-dimensional ${\cal N}=1$ field theories as well, where the fact that baryons are marginal can lead to large conformal manifolds \LeighEP\GreenDA.  Recently such theories were related to compactifications on Riemann surfaces of a certain minimal SCFT in six dimensions \RazamatGRO. It would be interesting to understand whether the constructions we consider here are useful in that context too.

\newsec{The model $T_x$}

We start by reviewing the known definitions of model $T_x$ and conjecturing a definition with both global symmetry and supersymmetry manifest in the UV theory.

\subsec{Description A: manifest global symmetry}

The first description has ${\cal N}=1$ supersymmetry and manifest $SU(3)$ global symmetry. This is simply a Wess--Zumino model of eight real superfields with superpotential,

\eqn\kfhgw{
d_{acb}\chi^a\chi^b\chi^c\,.
} Here $d_{abc}={\rm Tr} \,T_a\{T_b\,, T_c\}$ with $T_a$ the generators  of $SU(3)$. It was conjectured in \GaiottoYJH\BeniniBHK\ that the supersymmetry of this model enhances to ${\cal N}=2$ and a continuous R-symmetry emerges in the IR CFT.

\subsec{Description B: manifest supersymmetry}

\

\centerline{\figscale{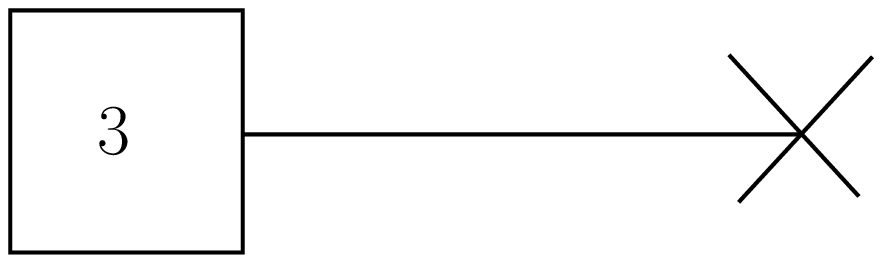}{2.3in}}
\medskip\centerline{\vbox{
\baselineskip12pt\advance\hsize by -1truein
\noindent\footnotefont{\bf Fig.~1:} We will denote the theory $T_x$ by such a graph and use it as a building block gauging either $SU(3)$ or $SU(3)/\Z_3$ symmetry with some Chern--Simons level.}}

\

\

A second description is the one which has emergent global symmetry but explicit supersymmetry. We will denote this description by $T_x$. Different descriptions might differ by contact terms and to be precise when referring to model $T_x$ we will refer to the model discussed in this section. The fact that we have manifest ${\cal N}=2$ supersymmetry allows us to utilize various localization techniques to study it \PestunZXK.

The  model  is an ${\cal N}=2$ supersymmetric  $U(1)$ gauge theory with two chiral fields with the same charge under the $U(1)$. We choose the charge to be one.
The global symmetry that we can identify in the Lagrangian is an $SU(2)$ rotating the two chiral fields, on top of which we have the topological $U(1)$ symmetry the monopole operators are charged under. 
It was claimed in \GaiottoYJH\BeniniBHK\ that the symmetry here enhances to $SU(3)$ with the fundamental given by the following decomposition into $SU(2)\times U(1)$,

\eqn\desfr{
{\bf 3} = {\bf 2}_{-1}+{\bf 1}_{2}\,.
} This can be easily seen by computing the supersymmetric index of the model. The index is the following measure of the spectrum of local operators:

\eqn\lsrgjhek{
{\cal I}= {\rm Tr}_{\S^2}\; \left[(-1)^{2J_3}q^{\frac12(\Delta+J_3)}\prod_a v_a^{e_a}\prod_b f_b^{q_b}\right]\,.
} Here $J_3$ is the generator of the $SU(2)$ rotation isometry of $\S^2$, $\Delta$ is the conformal dimension, $e_a$ are charges under the $a$-th Cartan generator of the global symmetry group, and finally the trace is taken in radial quantization. The fugacities $f_b$ are for abelian discrete symmetries $\Z_{n_b}$ and thus are $n_b$-th roots of unity, while $q_b$ label the elements of the discrete groups. 
 We will use in this paper the notations of the index of \AharonyKMA\ (which are explained in \AharonyDHA). For a review on the derivation of the index expressions the reader can consult \WillettADV. The index is a function of fugacities for different symmetries and fluxes for global symmetries through $\S^2$. 

Without turning on fluxes through the sphere for the global symmetry, the index is given by

\eqn\inderty{
{\cal I}(b,w)= \sum_{m=-\infty}^\infty w^{m} \oint\frac{dh}{2\pi i h} \,\underline{\cal I} (h b; m; r)\, \underline{\cal I} (h b^{-1};  m; r)\,.
} 
Here $h$ is the $U(1)$ gauge symmetry fugacity, $b$ that of the Cartan of $SU(2)$, $w$ of the topological $U(1)$, $\underline{\cal I}$ the index of a chiral field which is given by

\eqn\foief{
\underline{\cal I}(z;n;r) =\left(q^{\frac{1-r}2} z^{-1}\right)^{\frac{|n|}2} \prod_{l=0}^\infty\frac{1-(-1)^nz^{-1} q^{\frac{|n|}2+\frac12 r+l}}{1-(-1)^nz q^{\frac{|n|}2+1-\frac12 r+l}}\,.
} 
We take $r$ to be the R-charge.  The fugacity $z$ is for the $U(1)$ symmetry under which the chiral field is charged and $n$ is the flux through $\S^2$ for this symmetry. Note that the flux has to be properly quantized. The signs in the expressions appear as $J_3$. For an object of electric charge $e$ in the presence of a magnetic monopole with charge $m$, $J_3$ is shifted by $e\cdot m$ (see the discussion in \DimoftePY). 
The signs are important in general and we follow the notations of \AharonyKMA. The index of $T_x$  was analyzed in \GangWEK\ and here we will discuss some points which will be important for us. Evaluating the index one obtains,

\eqn\evaksie{
{\cal I}(b,w)=1-(2+b^2+b^{-2}+(b+b^{-1})(w+w^{-1}))q-(3+b^2+b^{-2}+(b+b^{-1})(w+w^{-1}))q^2+\cdots\,.
}
The term at order $q$ should count marginal operators minus conserved currents \BeemYN\RazamatGZX, and we see this is consistent with having no marginal operators and a current in the adjoint of $SU(3)$. The character of the adjoint is obtained upon taking $b=z_1^{-1/2} z_2^{-1}$ and $w = z_1^{-3/2}$ with $z_i$ parametrizing the Cartan of $SU(3)$ such that $\prod_{l=1}^3 z_l=1$.

For this theory the choice of R-symmetry for the chiral fields is a gauge symmetry and thus all choices should be equivalent. There is a small subtlety with this statement as we also have a Fayet--Iliopoulos parameter which is the mass for the topological symmetry. In presence of such a term, starting from some choice of R-symmetry and performing a gauge transformation that changes this assignment we produce a contact term between the R-symmetry and the topological one. Only for a particular choice of the contact term, for a given value of the R-symmetry, the theory will enjoy an $SU(3)$ symmetry. Let us exemplify this with the index computation.
The index in the presence of fluxes for the global symmetry is,

\eqn\kluwhfg{
{\cal I}(z_l; n, \hat n)=\left. \sum_{m \in \Z} w^{m} \oint\frac{dh}{2\pi i h}  h^{n }\, \underline{\cal I}  (h b; m+\hat n; r) \, \underline {\cal I} (h b^{-1};  m-\hat n; r)\right|_{w =z_1^{-3/2}, \, b=z_1^{-1/2} z_2^{-1}}\,.
} 

Here $n$ is the flux of the $U(1)$ topological symmetry and $\hat n$ that of the Cartan of the $SU(2)$ symmetry.  Note that changing the R-symmetry to $R\to R+2\alpha Q$ with $Q$ being charge under $U(1)$, amounts to redefining for the chiral fields $h\to q^\alpha h$. Without the FI parameter this has no effect on the index but with it we produce  a term of the form $q^{-\alpha n}$. This looks as a contact term between R-symmetry and the topological symmetry. 

The flux can be written, following the map of fugacities we derived here, in terms of the fluxes for the Cartan of $SU(3)$,

\eqn\efjl{
\left(n\, ,\; \hat n\right) = \left(-\frac32 m_1\, , \; -\frac12 m_1-m_2\right)\,.
} The index computed with arbitrary values of fluxes should be invariant under the action of the Weyl symmetry of $SU(3)$, that is

\eqn\dwkjfn{\eqalign{
&{\cal I}_x(z_i; m_i) := \left.{\cal I}(z_l; n, \hat n)\right|_{n=-\frac32 m_1,\; \hat n=-\frac12 m_1-m_2} \,, \;\;\;\;\;\;\;\cr
&{\cal I}_x(z_i; m_i)    = {\cal I}_x(z_{\sigma(i)}; m_{\sigma(i)})\, ,  \;\;\;\;\;\, \sigma \in S_3\,.}
} Here $m_3=-m_1-m_2$ and $z_3=z_1^{-1}  z_2^{-1}$. By computing the index we find that there is invariance if the R-charge is $1/3$ and there is no contact term. We can change the R-charge but then we will need to add a contact term between R-symmetry and the topological symmetry. 

Another interesting issue is the following subtlety. Note that, according to the way we define the theory, the magnetic monopole charges for $SU(2)$ and for the topological $U(1)$ are integers.  However, if the symmetry enhances to $SU(3)$ we should be able to turn on integer fluxes for the latter as well. According to \efjl, an odd $m_1$ flux would imply half-integer $n$ and $\hat n$ fluxes. To deal with this, whenever we turn on an odd $m_1$ we need to shift the lattice of fluxes for the gauge symmetry by a half.\foot{ See \RazamatPTA\ for a similar discussion in the context of ${\cal N}=4$ theories. Another way to phrase this is that the group rotating the chirals is $U(2)=(SU(2)\times U(1))/\Z_2$ and thus if we  gauge the $U(1)$ we can have half integer flux for $SU(2)$ as long as $U(1)$ has half integer flux.}
That is,

\eqn\klfejwhfg{
\eqalign{
&{\cal I}_x(z_i;m_i)=\cr
&\sum_{m \in \Z+\frac12 (m_1 \mod \, 2)} w^{m} \oint\frac{dh}{2\pi i h}  h^{-\frac32 m_1 } \underline{\cal I} \left(h b; m-\frac12 m_1-m_2; \frac13\right) \cdot\cr &\qquad \qquad \qquad \qquad \qquad \qquad \cdot \left. \underline {\cal I} \left(h b^{-1}; m+\frac12 m_1+m_2;\frac13\right)\right|_{w =z_1^{-3/2},\; b=z_1^{-1/2} z_2^{-1}}\,.}
} Computing the index, without refining with fugacities for the global symmetry, we obtain:

\eqn\lwerhwk{
{\cal I}_x(1;0) =1-8q-9q^2+18q^3+46q^4+\cdots\,.
}
Further, as we know from description A and as can be inferred from the index computation, the $N$-ality of all $SU(3)$ representations of states in the theory is $0$. This means that we can also turn on fluxes for $SU(3)$ which are shifted by multiples of $1/3$. The content of the model thus allows gauging of both $SU(3)$ and $SU(3)/\Z_3$.

\

\

\subsec{Description C: manifest global symmetry and supersymmetry}

We consider a Wess--Zumino model with nine ${\cal N}=2$ chiral  superfields organized into a bi-fundamental chiral $Q_{ij}$ of two $SU(3)$ symmetries, and a superpotential given by the baryon\foot{Such theories in three dimensions flow to interacting SCFTs in three dimensions, {\it e.g.} they have intricate conformal manifolds, see  for example \BaggioMAS.}

\eqn\syeur{
W=\epsilon^{ilm}\epsilon^{jkc}Q_{ij}Q_{lc}Q_{mk}\,.
} We then gauge one of the $SU(3)$ symmetries with a level-$5/2$ Chern--Simons term. For smaller values of the level the theory will be ``bad'', that is the partition function will not be well-defined. 
This might signal either spontaneous breakdown of supersymmetry, or wrong R-symmetry assignments as in \GaiottoAK. Moreover we need half-integer Chern--Simons level, for otherwise the theory would have a parity anomaly. 

The model has manifest $SU(3)$ symmetry and ${\cal N}=2$ supersymmetry and we conjecture it is dual to $T_x$.  As a check one can compute the supersymmetric index and find that at least in expansion in fugacities it matches precisely with the index of $T_x$.  Importantly, the two models have a relative contact Chern--Simons term for the global $SU(3)$ symmetry at level one. To see this we can compute the index in presence of a background monopole flux for the $SU(3)$ symmetry. For example, taking the flux to be $(m_1,m_2,m_3)=(1,-1,0)$ we obtain, for the model $T_x$:

\eqn\lfjhe{
q^{1/2}\left(z_1^{-1/2} z_2^{-5/2}+z_1^{1/2}z_2^{5/2}\right)-q \left(z_1^{1/2}z_2^{-1/2}+z_2^{1/2}z_1^{-1/2} \right)+\cdots\,.
} 
For the $SU(3)$ gauge theory we obtain instead:

\eqn\jhfg{
q^{1/2}\left(z_1^{1/2} z_2^{-7/2}+z_1^{3/2}z_2^{3/2}\right)-q \left(z_1^{1/2}z_2^{-1/2}+z_2^{-3/2}z_1^{3/2}\right)+\cdots\,.
}
Since the Chern--Simons term contributes $z_1^{k(2m_1+m_2)}z_2^{k(2m_2+m_1)}$ to the index, where $(m_1,m_2,-m_1-m_2)$ is the flux and $k$ the level, we can see that the two above expressions differ by a factor of $z_1 z_2^{-1}$, which comes from a background Chern--Simons term at level one. We stress that this model has manifest symmetry and supersymmetry to be contrasted with the other descriptions.\foot{However, because of the Chern-Simons term this description does not have manifest time reversal symmetry with the two other descriptions manifestly invariant. We thank Kazuya Yonekura for pointing this fact out to us.}

\centerline{\figscale{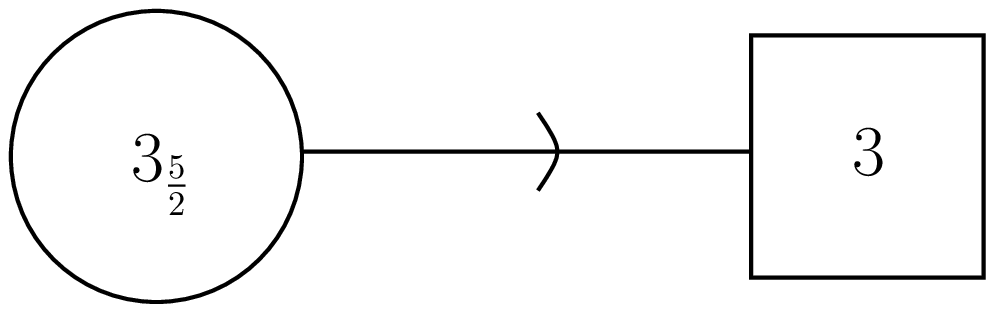}{2.3in}}
\medskip\centerline{\vbox{
\baselineskip12pt\advance\hsize by -1truein
\noindent\footnotefont{\bf Fig.~2:} The theory $T_x^{(2)}$ which is dual to $T_x$ with a contact term for the $SU(3)$ global symmetry.  We turn on a baryonic superpotential for the chiral field which preserves the non-abelian symmetry. Here we parametrize the gauge node by $N$ of $SU(N)$ and by the level $k$ of the Chern--Simons term as $N_k$.}} 

\

The $SU(N)$ gauge theories with matter in fundamental representations in three dimensions have known duals which descend from dualities in four dimensions and real mass deformations thereof \AharonyDHA\ParkWTA. One can use such dualities to obtain other descriptions with manifest symmetry. For example, following \BeniniMF\ (see also \debult), we know that ${\cal N}=2$ $U(3)$ at level $5/2$ with three fundamental chirals is dual to a $U(1)$ gauge theory at level $-5/2$ with three fundamental chirals. Importantly there are also contact terms, one of which is a Chern--Simons term at level one for the topological symmetry. We can gauge the topological symmetry of the pair by putting the Chern--Simons term for the topological symmetry on the $U(1)$ side. The $U(3)$ model then becomes the $SU(3)$ theory we consider \KapustinSim, and on the dual side (because of the contact term) we are left with a $U(1)$ gauge theory. The baryonic symmetry of the $SU(3)$ theory maps to the topological symmetry on the dual side; turning on a baryonic superpotential amounts to a monopole superpotential on the $U(1)$ side.\foot{Monopole superpotentials were first discussed in the context of dualities in \AharonyGP. See \BenvenutiWET\BeniniDUD\AmaritiGDC\GiacomelliVGK~for many recent examples of models with monopole superpotentials.} The reader can consult \AharonyUYA\ for this duality and we work out the details in the Appendix. We moreover use it to verify the equality of $\S^3$ partition functions of the new description and description B. This is an independent check of the duality.

Let us note that we can consider a generalization of the model by increasing the Chern--Simons level, though we do not have any claim for duality for higher levels.
We consider gauging with arbitrary CS term at level $k= l+\frac12$, with integer $l$ bigger than one. All such models have $SU(3)$ global symmetries and no ${\cal N}=2$ marginal or relevant deformations. We will denote such models as $T_x^{(l)}$, so that $T^{(2)}_x$ is dual to $T_x$ (adding contact terms). Increasing $l$ the monopoles will obtain higher charges meaning the gauge sector will have weaker coupling.  The index for several values of $l$ is:

\eqn\ldjfghkw{\eqalign{
&l=2 \,:\;\;\;\, 1-8q-9q^2+18q^3+46 q^4+\cdots\,\;; \cr
&l=3 \,:\;\;\;\, 1-8q+9q^2+53q^3+28 q^4+\cdots\,\;;\cr
&l=4 \,:\;\;\;\, 1-8q+9q^2+43q^3-9 q^4+\cdots\,\;.
}
}The $-8$ at order $q$ is the contribution of the conserved currents of the $SU(3)$ global symmetry. For higher values of the Chern--Simons level the first terms in the expansion of the index are as for $l=4$ and the difference appears at higher powers of $q$, as it comes from gauge-invariant dressed monopole operators (whose charges scale with the level).

\

\

\newsec{Gluing the $T_x$ together and ${\cal N}=2$ mirror duality}

We can consider gluing together several copies of $T_x$ by gauging the diagonal $SU(3)$ or $SU(3)/\Z_3$ symmetry with a Chern--Simons term. As the model $T_x$ does not have any marginal or relevant deformations for large enough values of the level of the Chern--Simons term or large enough number of copies, the resulting models will also have no relevant or marginal deformations as the charges of the monopoles of the gauge group increase with the level and the amount of matter. For too low a number of copies and too low a level, the theories are bad in the sense that the partition functions do not converge.  However, for high enough levels and number of copies the theories are sensible and might have interesting low dimension operators. We have studied the models with minimal levels and number of copies possible and in what follows
we will report on  few examples where we could recognize a dual description. We find  three  examples with the theories  dual to simple ${\cal N}=2$ Wess--Zumino models with cubic interactions.\foot{In \GangWEK, motivated by geometric considerations, the authors considered gauging subgroups of $SU(3)$ for a single $T_x$.} We will moreover discuss a dual of $T_x$ itself which can be obtained via such a construction. The main check we will refer to is the equality of indices. We have verified such equalities in a series expansion in $q$ to several non-trivial orders, but do not have a proof of the identities.

\

 \
 
 \centerline{\figscale{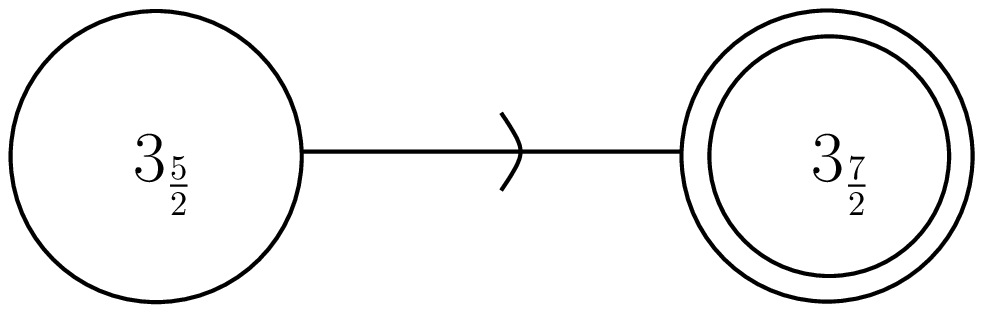}{2.3in}}
\medskip\centerline{\vbox{
\baselineskip12pt\advance\hsize by -1truein
\noindent\footnotefont{\bf Fig.~3:} The dual of theory  $T_x$ with $SU(3)/\Z_3$ gauged at level $k=7/2$. The theory is dual to a chiral field with cubic superpotential. The double circle denotes $SU(3)/\Z_3$ gauging with the single circle denoting an $SU(3)$ gauge model.}}

\

\subsec{Single $T_x$ with $SU(3)/\Z_3$ gauged dual to WZ with one chiral field}

Let us consider gauging the $SU(3)/\Z_3$ symmetry of a single copy of $T_x$. The model then will have no continuous global symmetry and will have $\Z_3$ symmetry. With low level of Chern--Simons term the theory is bad, and the lowest level for which we find that the partition function converges is $9/2$. We also find that the index of this model agrees with the index of a single chiral field with cubic superpotential. We conjecture then that $T_x$ with $SU(3)$ gauged at level $9/2$ is a Wess--Zumino model of one chiral field with cubic superpotential.
The index is given by,\foot{For a discussion of gauging of $SU(N)/\Z_N$ symmetries in the index see e.g. \RazamatPTA. The index in three dimensions can be obtained as the limit of the lens index in four dimensions \BeniniNC\ which depends in a non-trivial way on the global structure of the gauge group \RazamatOPA. The differences in monopole operators appearing for different global structures are relatives of differences in line operators, see \AharonyHDA\   for a recent discussion in four dimensions.}
\eqn\kdjfh{\eqalign{
&\frac16\sum_{l=0}^2 g^l\sum_{m_1,m_2\in \Z+\frac{l}3}\oint \frac{dz_1}{2\pi i z_1} \frac{dz_2}{2\pi i z_2}   \prod_{i\neq j} q^{-\frac{|m_i-m_j|}4} \left(1-(-1)^{m_i-m_j} q^{\frac{|m_i-m_j|}2} \frac{z_i}{z_j}\right) \cdot \cr&\;\;\;\;\;\,  \cdot (z_1^{2m_1+m_2}z_2^{2m_2+m_1})^{9/2} \, {\cal I}_x(z_i; m_i)=\underline{\cal I}\left(g^{-1};0;\frac23\right)\,.
}
} Here $g$ is a third root of unity, i.e. a fugacity for the $\Z_3$ symmetry. The operators charged under the discrete symmetry are gauge-invariant monopole operators with fractional charge.\foot {Weighting different sectors by the discrete symmetry is a three-dimensional avatar of weighting different sectors in the lens index in four dimensions \RazamatOPA.}
The theory with gauged $SU(3)$ symmetry is then dual to the $\Z_3$ gauging of the Wess--Zumino model, i.e.

\eqn\kgfj{\eqalign{
&\frac16 \sum_{m_1,m_2\in \Z}\oint \frac{dz_1}{2\pi i z_1}\frac{dz_2}{2\pi i z_2}     \prod_{i\neq j} q^{-\frac{|m_i-m_j|}4} \left(1-(-1)^{m_i-m_j} q^{\frac{|m_i-m_j|}2} \frac{z_i}{z_j}\right) \cdot \cr&\;\;\;\;\;\,  
  \cdot (z_1^{2m_1+m_2}z_2^{2m_2+m_1})^{9/2} \, {\cal I}_x(z_i; m_i)=\frac13\sum_{j=0}^2\underline{\cal I}\left(e^{\frac{2\pi i j}3};0;\frac23\right)\,.
}
} 
We can use description C to write a quiver (see Figure 2), and as descriptions B and C have a relative Chern--Simons contact term for the $SU(3)$ symmetry, the level of the $SU(3)/\Z_3$ Chern--Simons term we need to add is $7/2$. 
 
 Note that the mirror dual of a free chiral field is well-known, and is given by a $U(1)$ gauge theory with level half Chern--Simons term plus a single chiral field. The topological symmetry is dual to the $U(1)$ baryonic symmetry rotating the chiral field. We need to turn on a cubic  superpotential to break this symmetry, which on the gauge theory side translates to a monopole superpotential. Thus this provides yet another dual of the model we build by gauging the $SU(3)/\Z_3$ symmetry of $T_x$.
 
 \
 
 \

\centerline{\figscale{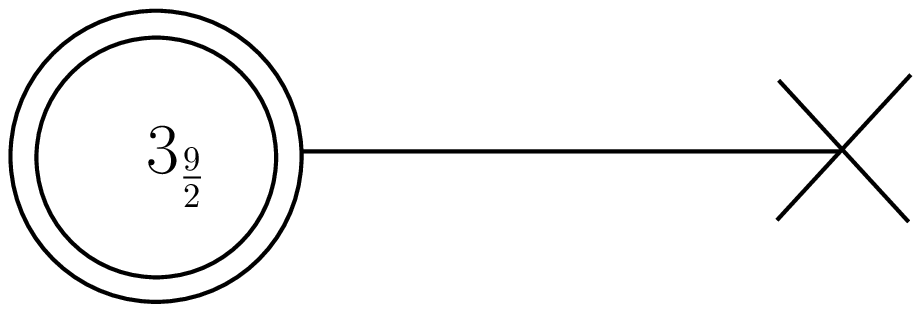}{2.3in}}
\medskip\centerline{\vbox{
\baselineskip12pt\advance\hsize by -1truein
\noindent\footnotefont{\bf Fig.~4:} The theory  $T_x$ with $SU(3)/\Z_3$ gauged at level $k=9/2$. The theory is dual to a chiral field with cubic superpotential. The double circle denotes the $SU(3)/\Z_3$ gauging. The different level here and in the previous figure is because the duality involves a relative contact term. }}

\

\subsec{Five glued $T_x$ dual to WZ with five chiral fields}

We consider gauging a diagonal $SU(3)/\Z_3$ symmetry of five copies of $T_x$ at level $3/2$. The index is equal to the one of the Wess--Zumino model with five chiral fields and a general cubic superpotential,

\eqn\fekgf{
W= \sum_{i,j,l =1}^5 \lambda_{ijl} \Phi_i \Phi_j \Phi_l\,.
} The index is

\eqn\kdjfh{\eqalign{
&\frac16\sum_{l=0}^2 g^l \sum_{m_1,m_2\in \Z+\frac{l}3}\oint \frac{dz_1}{2\pi i z_1} \frac{dz_2}{2\pi i z_2}   \prod_{i\neq j} q^{-\frac{|m_i-m_j|}4} \left(1-(-1)^{m_i-m_j} q^{\frac{|m_i-m_j|}2} \frac{z_i}{z_j}\right) \cdot \cr&\;\;\;\;\;\,  \cdot (z_1^{2m_1+m_2}z_2^{2m_2+m_1})^{\frac32}  {\cal I}_x(z_i; m_i)^5=\underline{\cal I}\left(g^{-1};0;\frac23\right)^5\,.
}
} This suggests that gauging five copies of $T_x$ with $SU(3)/\Z_3$ at level $3/2$ is dual to a Wess--Zumino model of five chiral fields. The model has a $\Z_3$ global symmetry which we can gauge.

\

\

\centerline{\figscale{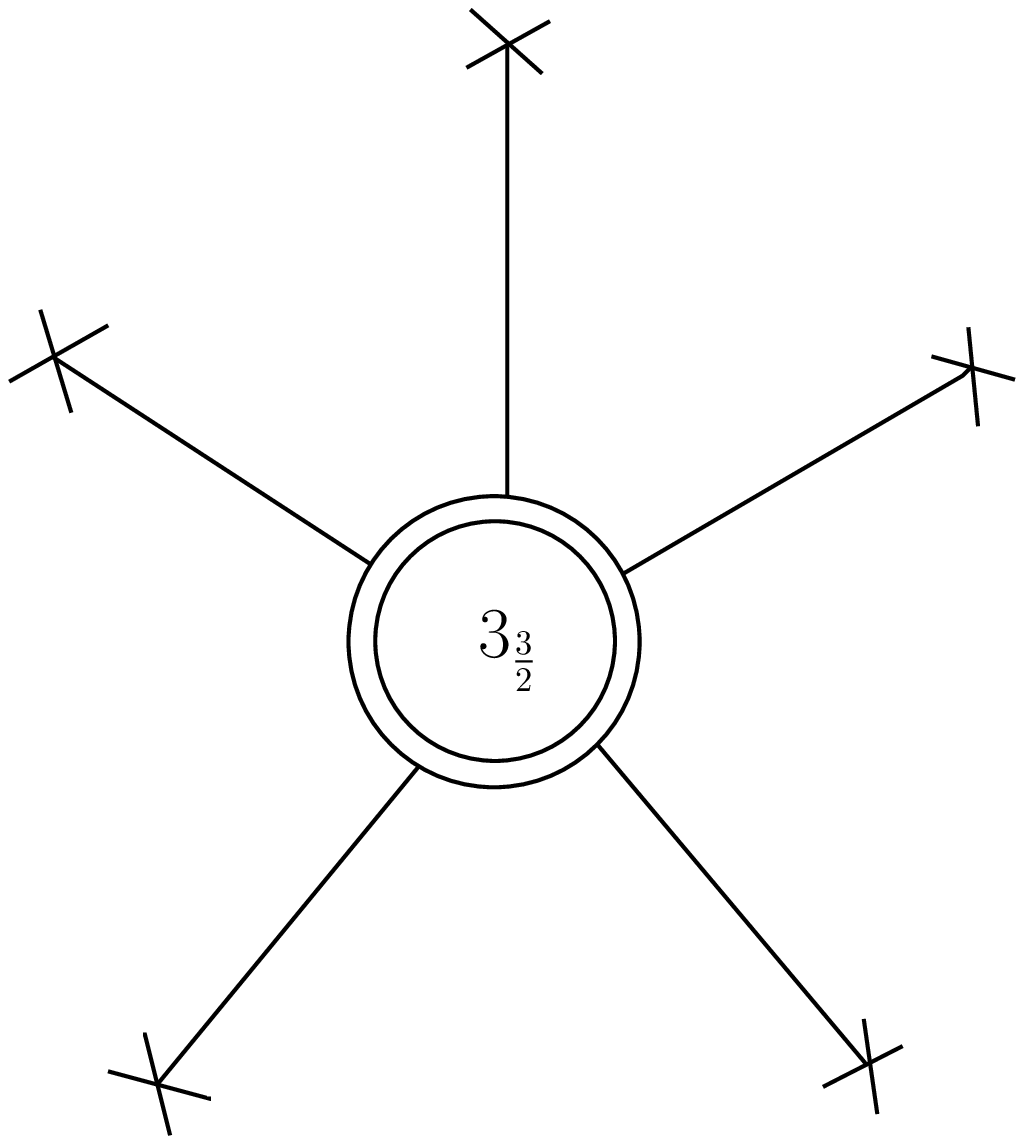}{2.3in}}
\medskip\centerline{\vbox{
\baselineskip12pt\advance\hsize by -1truein
\noindent\footnotefont{\bf Fig.~5:} The theory of five copies of $T_x$ with $SU(3)/\Z_3$ gauged at level $k=3/2$, dual to a WZ model of five chiral fields.}} 

\

\subsec{Eight glued $T_x$ dual to WZ with fourteen chiral fields}

We consider gauging a diagonal $SU(3)/\Z_3$ symmetry of eight copies of $T_x$ at level zero. The index is equal to the one of a Wess--Zumino model with fourteen chiral fields and a general cubic superpotential,

\eqn\fekgf{
W= \sum_{i,j,l =1}^{14} \lambda_{ijl} \Phi_i \Phi_j \Phi_l\,.
} The index is

\eqn\kdjfh{\eqalign{
&\frac16\sum_{l=0}^2 g^l \sum_{m_1,m_2\in \Z+\frac{l}3}\oint \frac{dz_1}{2\pi i z_1} \frac{dz_2}{2\pi i z_2}   \prod_{i\neq j} q^{-\frac{|m_i-m_j|}4} \cdot \cr &\;\;\;\;\;\;\,\, \cdot \left(1-(-1)^{m_i-m_j} q^{\frac{|m_i-m_j|}2} \frac{z_i}{z_j}\right) \; {\cal I}_x(z_i; m_i)^8=\underline{\cal I}\left(g^{-1};0;\frac23\right)^{14}\,.}
} This suggests that gauging eight copies of $T_x$ with $SU(3)/\Z_3$ at level zero is dual to a Wess--Zumino model of fourteen chiral fields. 

\centerline{\figscale{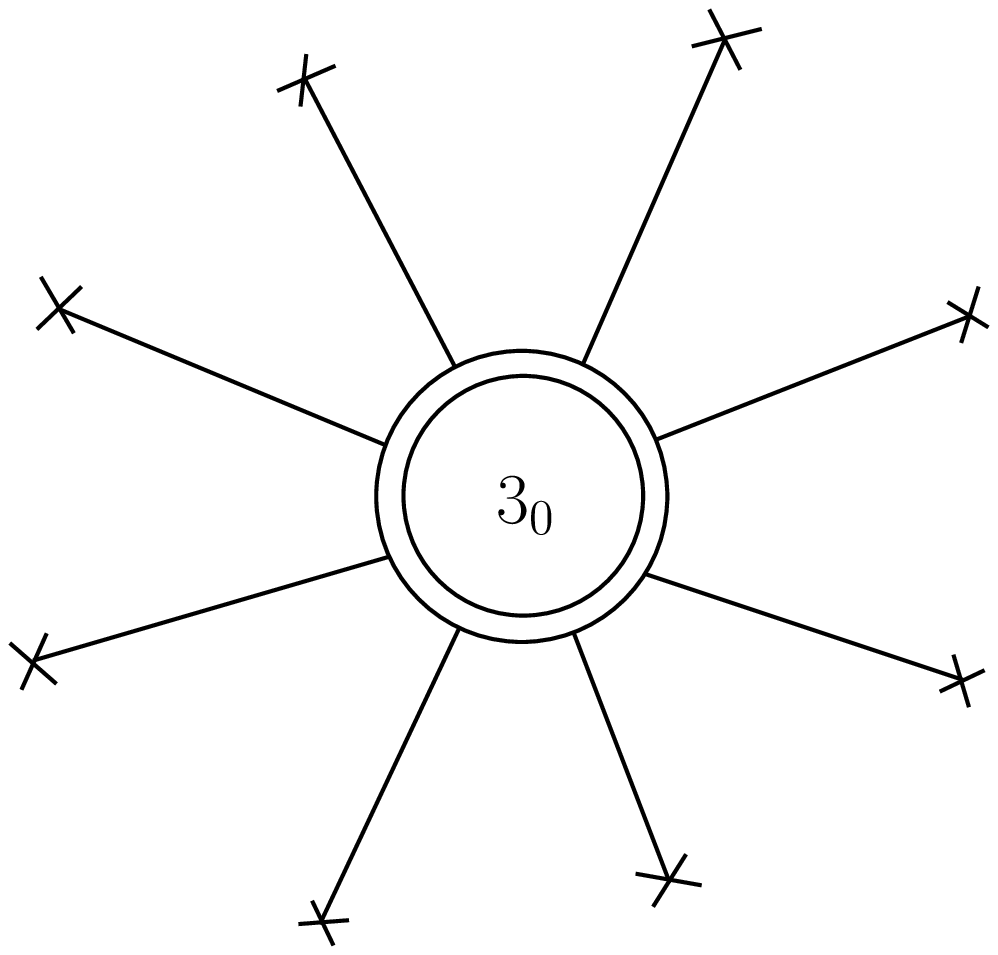}{2.3in}}
\medskip\centerline{\vbox{
\baselineskip12pt\advance\hsize by -1truein
\noindent\footnotefont{\bf Fig.~6:} The theory of eight copies of $T_x$  with $SU(3)/\Z_3$ gauged at level zero, dual to a WZ model with fourteen chiral fields.}} 

In this case the $\S^3$ partition function is converging fast enough, so we can evaluate it in both dual frames. The $\S^3$ partition function of $T_x$ is

\eqn\elfhg{
Z_x(m_1,m_2)=\int_{-\infty}^\infty d\sigma e^{2\pi i \sigma(-\frac32 m_1)} e^{l(\frac23+i \sigma-i \frac12 m_1-i m_2)+l(\frac23+i\sigma +i \frac12 m_1+i m_2)}\,.
} Here $l(m)$ is the $l$-function of Jafferis \JafferisUN\ and $m_i$ are real masses for the $SU(3)$ symmetry.  The duality implies that

\eqn\wrkhfgwk{
\frac12 \int_{-\infty}^\infty d\sigma_1\int_{-\infty}^\infty d\sigma_2 \prod_{i\neq j}|2\sinh\pi(\sigma_i-\sigma_j)| Z_x(\sigma_1,\sigma_2)^8=\left[e^{l(\frac13)}\right]^{14}\,.
} Note that the half in front of the integral is $1/3!$, the dimension of the Weyl group, times $3$, coming from the fact that the gauging is of $SU(3)/\Z_3$. (See \BenvenutiGA\  for similar factors in class ${\cal S}$ computations.) 
We find that the equality indeed holds, and the numerical evaluation yields $0.01706$ for the first five digits. This is an independent check of the duality.   

For eight and more copies of $T_x$ glued there is no need for Chern--Simons terms for the partition functions to converge; thus we can assume that these describe SCFT's and we identified the R-symmetry correctly. The index of these models with $s$ copies of $T_x$ and $SU(3)/\Z_3$ gauged is given by, for $s=8,10,12,14,16$,

\eqn\krghik{
1+C_{\frac{s}{2}} q^{\frac{s/2-3}{3}}+\cdots\,.
} Here $C_n$ is the $n$-th Catalan number, $(2n)!/((n+1)! n!)$. The operators contributing to the leading order are gauge-invariant dressed monopole operators. As we have just seen, in the $s=8$ case we have a dual description where the basic operators come from chiral fields and it is interesting to understand whether there are duals for higher values of $s$ such that the basic operators do not come from monopoles, a question we leave for future investigation.
 For higher values of $s$ the index starts with $q^2$. We can therefore deduce that there are states associated with monopole operators which are counted by the Catalan number, and at $q^2$ other states appear. The number of new states at $q^2$, which for high enough $s$ is all the states, is $s(s-3)/2$.
 
 \

\centerline{\figscale{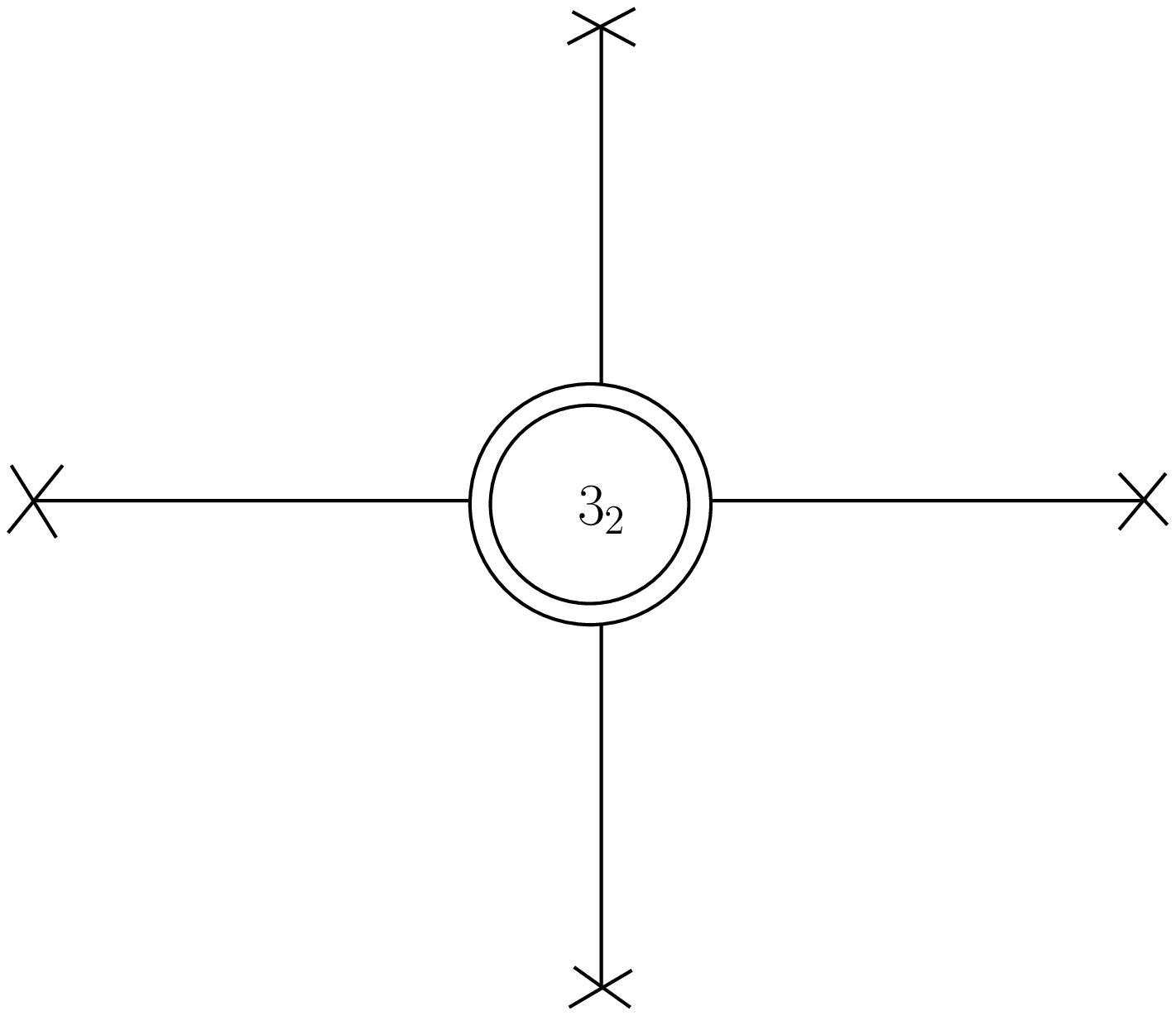}{2.3in}}
\medskip\centerline{\vbox{
\baselineskip12pt\advance\hsize by -1truein
\noindent\footnotefont{\bf Fig.~7:} Model of four copies of $T_x$ with $SU(3)/\Z_3$ gauged at level two, dual to $T_x$ itself.}} 

\

\subsec{Four glued $T_x$ dual to $T_x$}

We consider gauging a diagonal $SU(3)/\Z_3$ symmetry of four copies of $T_x$ at level two. The index is equal to the (unrefined) index of a single $T_x$:
\eqn\kdjfh{\eqalign{
&\frac16\sum_{l=0}^2 g^l \sum_{m_1,m_2\in \Z+\frac{l}3}\oint \frac{dz_1}{2\pi i z_1} \frac{dz_2}{2\pi i z_2}   \prod_{i\neq j} q^{-\frac{|m_i-m_j|}4}\left(1-(-1)^{m_i-m_j} q^{\frac{|m_i-m_j|}2}\frac{z_i}{z_j}\right) \cdot \cr&\;\;\;\;\;\,  \cdot (z_1^{2m_1+m_2}z_2^{2m_2+m_1})^{2}  \; {\cal I}_x(z_i; m_i)^4={\cal I}_x(1;0)\,.
}
} Note that this turns out to be independent of $g$, and the same as the index for a gauged $SU(3)$. This suggests that gauging four copies of $T_x$ with $SU(3)/\Z_3$ at level $2$ is dual to $T_x$. The model has a $\Z_3$ symmetry which is identified with the center of the $SU(3)$ symmetry of $T_x$. Given that only representations with zero $N$-ality appear, we do not observe it in the  computation.
 
\

\

\

\noindent{\it Acknowledgments: } We would like to thank Antonio Amariti, Sergio Benvenuti, Zohar Komargodski, Brian Willett, Kazuya Yonekura, and Gabi Zafrir for comments and relevant discussions. This research was supported by the Israel Science Foundation under grant No. 1696/15 and by the I-CORE Program of the Planning and Budgeting Committee. The research of MF was also supported by the Israel Science Foundation under grant No. 504/13.

\

\appendix{A}{$U(1)$  SQED with monopole superpotential dual of $T_x$ with manifest symmetry and supersymmetry}

We start with a $U(1)$ theory at Chern--Simons level $-5/2$ and three chiral fields with charge one. We give R-charge zero to the chiral fields for concreteness. The theory has $SU(3)$ symmetry rotating the fields and a topological $U(1)$. 
 According to \BeniniMF\ this model is dual to a $U(3)$ gauge theory with three fundamental flavors and level $5/2$ Chern--Simons term. The R-charge of the fields is one. We also have a level-one relative contact term for the topological symmetry, a level-one relative Chern--Simons term for $SU(3)$, and a level-$3/2$ relative mixed Chern--Simons term for the topological $U(1)$ and the R-symmetry.  We put all the contact terms on the $U(1)$ side of the duality; the duality implies e.g. that the index of the $U(3)_{5/2}$ theory with three fundamental fields is equal to
 
 \eqn\ekhf{
 w^{-n}q^{-\frac34\;n}\sum_{m\in \Z} w^m \oint\frac{dz}{2\pi i z} z^n z^{-\frac52 m} \prod_{j=1}^3\underline {\cal I}(z z_j; m; 0)\,.
 }  We have turned on a magnetic flux $n$ for the topological symmetry.
 Now we can gauge the topological $U(1)$ symmetry. Since on the $U(3)$ side we do not have any contact terms, we will obtain an $SU(3)$ gauge theory at level $5/2$ and three fundamental chiral fields. On the dual side we perform the analysis using the index. The gauging of the topological symmetry gives
 
 \eqn\efkjh{\eqalign{
 &\sum_{n\in \Z} c^n \oint \frac{dw}{2\pi i w}\left[w^{-n}q^{-\frac34\;n}\sum_{m\in \Z} w^m \oint\frac{dz}{2\pi i z} z^n z^{-\frac52 m} \prod_{j=1}^3\underline {\cal I}(z z_j; m; 0)\right] \cr
 &=\sum_{n\in \Z}  c^n q^{-\frac34\;n} \oint\frac{dz}{2\pi i z}  z^{-\frac32 n} \prod_{j=1}^3\underline {\cal I}(z z_j; n;0)\,.
 }
 } Here we performed  the integral over $w$ which identified the flux of the original gauge symmetry $m$ with the flux of the topological symmetry $n$. The fugacity $c$ is for the topological symmetry of the new $U(1)$ gauge symmetry. This is dual to $-1/3$ the baryonic symmetry on the $SU(3)$ gauge theory side. We can evaluate this index to be
 
 \eqn\ywg{
 1-9q+(c^{-1} -10 c)q^{\frac32}-18 c q^{\frac52}+(44+c^{-2})q^3+\cdots\,.
 } Here the $9$ at order $q$ is the conserved current for the baryonic $U(1)$ and for the $SU(3)$. The term with weight $c^{-1}$ at order $q^{\frac32}$ is the baryon on the $SU(3)$ side. Remember that the R-charge of the quarks on the $SU(3)$ side is one and thus this is precisely how the baryon contributes. On the $U(1)$ side as this state is charged under the topological symmetry it comes from the monopoles.
  Now we need to turn on the baryonic superpotential. This amounts on the $U(1)$ side to a monopole superpotential. In the index we need to set $c^{-1}q^{\frac32}$ to $q$. The index then becomes,
  
  \eqn\kjrgf{
\sum_{n\in \Z}   q^{-\frac14n} \oint\frac{dz}{2\pi i z}  z^{-\frac32 n} \prod_{j=1}^3\underline {\cal I}(z z_j; n; 0) = \sum_{n\in \Z} \oint\frac{dz}{2\pi i z}  z^{-\frac32 n} \prod_{j=1}^3\underline {\cal I}\left(z z_j; n;-\frac13\right)
  \,.
} This is just the index of a $U(1)$ gauge theory at level $-3/2$ plus three charge-one fields with R-charge $-1/3$, and a monopole superpotential. The claim is that the latter is dual to an $SU(3)$ model at level $5/2$ 
with three fundamental fields with R-charge one. This is nothing but $T_x$ up to contact terms. 
The index above is

\eqn\ekhjfgek{
1-8q-9q^2+18q^3+46q^4+\cdots\,.  \;\;\,\;
} This agrees with the computations done in the $T_x$ model in an expansion in fugacities. Computing the index, refined with the fugacities and magnetic fluxes for the global symmetry, we can deduce that the model discussed here has a level minus one contact term for the $SU(3)$ symmetry relative to $T_x$. We can refine the index with fluxes for $SU(3)$ and discover that there is a relative Chern--Simons contact term for that symmetry at level one. 

The $U(1)$ description with the monopole superpotential is simple enough to allow for a numerical evaluation of  the $\S^3$ partition function. We have checked extensively that the latter agrees with that of $T_x$ as a function of the real mass parameters for the $SU(3)$ flavor symmetry. The precise equality is

\eqn\wkurhfi{
Z_x(m_1, m_2)= e^{2\pi i (m_1^2+m_2^2+m_1m_2)}e^{\frac{\pi i 19}{12}}\int_{-\infty}^{\infty}d\sigma e^{-\frac32\pi i (\sigma+i\epsilon)^2}e^{\pi (\sigma+i\epsilon)} e^{\sum_{j=1}^3l(1+i(\sigma+i \epsilon+m_j))}\,.
} Here $Z_x$ is the partition function of $T_x$ which was given in \elfhg. The parameter $\epsilon$ is an arbitrarily small positive real number which lifts the contour of integration slightly above the real axis in the complex plane. This is necessary to avoid poles whenever $\sigma=-m_i$. The parameters $m_i$ are real masses for $SU(3)$ and satisfy $\sum_{l=1}^3 m_l =0$. The two exponentials in front of the integral are contact terms, with the first being at level one for the $SU(3)$ symmetry and second for the R-symmetry.  We also stress that the equality of the partition function of the description we discuss here and description C is a mathematical identity following from \debult.\foot{In fact the equality above itself follows from Eq. (5.6.18) in \debult, which was given a physical interpretation in \amacassi\ (see also \benve).} Therefore checking its equality with $T_x$ is equivalent to checking the equality of description C with $T_x$.

\listrefs
\end